\documentclass{aa}  

\usepackage{graphicx}
\usepackage{txfonts}
\newcommand{\be}{\begin{equation}}
\newcommand{\ee}{\end{equation}}

\begin{document}

   \title{Accurate modelling of the Lyman-$\alpha$ coupling for the 21-cm signal, observability with NenuFAR and SKA}

   \author{B. Semelin\inst{1}\fnmsep\thanks{benoit.semelin@obspm.fr}
           \and
           R. Mériot\inst{1}
           \and
           F. Mertens\inst{1}\fnmsep\inst{2}
           \and 
           L. V. E. Koopmans \inst{2}
           \and
           D. Aubert \inst{3}
           \and
           R. Barkana \inst{4,5,6}
           \and
           A. Fialkov \inst{7,8}
           \and
           S. Munshi \inst{2}
           \and
           P. Ocvirk \inst{3}
         }

   \institute{Sorbonne Université, Observatoire de Paris, PSL research university, CNRS, LERMA, F-75014 Paris, France\\
             \and 
             Kapteyn Astronomical Institute, University of Groningen, PO Box 800, 9700AV Groningen, The Netherlands\\
             \and
             Observatoire Astronomique de Strasbourg, Université de Strasbourg, CNRS UMR 7550, 11 Rue de l’Université, F-67000 Strasbourg, France\\
             \and
             School of Physics and Astronomy, Tel Aviv University, Tel Aviv, 69978, Israel\\
             \and
             Institute for Advanced Study, 1 Einstein Drive, Princeton, New Jersey 08540, USA \\
             \and
             Department of Astronomy and Astrophysics, University of California, Santa Cruz, CA 95064, USA \\
             \and
             Institute of Astronomy, University of Cambridge, Madingley Road, Cambridge, CB3 0HA, UK\\
             \and
             Kavli Institute for Cosmology, Madingley Road, Cambridge, CB3 0HA, UK}

   \date{Received X X, 2022; accepted X X, 2022}

 
\abstract{The measurement of the $21$ cm signal from the Cosmic Dawn is a major goal for several existing and upcoming radio interferometers such as NenuFAR and the SKA. During this era before the beginning of the Epoch of Reionization, the signal is  more difficult to observe due to brighter foregrounds but reveals additional information on the underlying astrophysical processes encoded in the spatial fluctuations of the spin temperature of hydrogen. To interpret future measurements, controlling the level of accuracy of the Lyman-$\alpha$ flux modelling is mandatory. In this work, we evaluate the impact of various approximations that exist in the main fast modelling approach compared to the results of a costly full radiative transfer simulation. The fast SPINTER code, presented in this work, computes the Lyman-$\alpha$ flux including the effect of wing scatterings for an inhomogeneous emissivity field, but assuming an otherwise homogeneous expanding universe. The LICORICE code computes the full radiative transfer in the Lyman-$\alpha$ line without any substantial approximation. We find that the difference between homogeneous and inhomogeneous gas density and temperature is very small for the computed flux. On the contrary, neglecting the effect of gas velocities produces a significant change in the computed flux. We identify the causes (mainly Doppler shifts due to velocity gradients) and quantify the magnitude of the effect in both an idealised setup and a realistic cosmological situation. We find that the amplitude of the effect, up to a factor of $\sim 2$ on the $21$ cm signal power spectrum on some scales (depending on both other model parameters and the redshift), can be easily discriminated with an SKA-like survey and already be approached, particularly for exotic signals, by the ongoing NenuFAR Cosmic Dawn Key Science Program.   }

\maketitle
%

\section{Introduction}

The Cosmic Dawn (CD) is the period at the beginning of the Epoch of Reionisation (EoR) when the first stars formed. A more quantitative definition, born from the study of the 21-cm signal emitted by the intergalactic medium (IGM) during the EoR, is to say that the Cosmic Dawn  corresponds to the period when the fluctuations of the $21$-cm signal were dominated not by fluctuations of the ionisation or density fields but rather by fluctuations of the spin temperature of hydrogen, which were in turn regulated by fluctuations of the gas kinetic temperature and of the strength of the Wouthuysen-Field coupling by Lyman-$\alpha$ photons \citep{Wouthuysen52,Field58}. In practice, for the more standard models, the CD corresponds to an averaged ionisation fraction of the IGM of less than a few percent. Note that the corresponding redshift range is model-dependent: the CD occurs earlier if low-mass halos are able to form stars efficiently and later if not. If, as can be expected, atomic cooling halos (halos able  to cool below their virial temperature through atomic cooling only, that is halos with mass $\gtrsim 10^8$ M$_\odot$) do form stars efficiently, the CD occurs at $z \gtrsim 15$, meaning that the signal must be observed at frequencies $\lesssim 90$ MHz. In this regime, the signal will be seen in absorption against the CMB and will exhibit brightness temperature fluctuations on large scales with an amplitude up to several tens of mK.  

So called global experiments attempt to detect the signal averaged over the whole sky through its frequency dependence only. Separation from foregrounds and possible instrumental effects is then an especially challenging task considering the limited leverage offered by the available information. Consequently, at this stage, such experiments report incompatible results: the EDGES experiment claims a detection around $78$ MHz \citep{Bowman18} while the SARAS 3 experiment excludes the same signal with a $95 \%$ confidence level \citep{Singh22}. Interferometers, on the other hand, have the ability to measure angular fluctuations of the signal and thus, in principle, produce a full tomography of the signal . This, however, requires very high sensitivity and not even the SKA will be able to produce tomographic images during the Cosmic Dawn \citep[see, e.g.][]{Mellema13,Koopmans15}. The three-dimensional isotropic power spectrum benefits from a better signal-to-noise ratio while retaining more information than the global signal. Nevertheless, published upper limits on current instruments able to probe the CD (LWA, MWA, LOFAR, AARTFAAC) are orders of magnitude above the expected level of the signal \citep{Ewall-Wice16,Gehlot19,Eastwood19,Gehlot20,Yoshiura21}. In this work, we will compare our modelled signals to the expected sensitivity of NenuFAR \citep{Mertens21} and SKA.

From the upper limits or a detection of the signal, the parameters of astrophysical models can be inferred using either a classical MCMC approach \citep[e.g.][]{Greig15,Greig17} or with methods involving some aspects of machine learning \citep{Shimabukuro17,Gillet19,Schmit18,Jennings19,Doussot19,Cohen20,Hortua20,Bevins22,Zhao22,Bye22,Abdurashidova22}. In all cases, the modelling of the signal is a fundamental step of the inference process: either at each step of the MCMC approach or for building a learning sample in supervised learning based methods. The maximum likelihood values and posterior distribution of the astrophysical parameters will be affected by the approximations made in the modelling step. The modelling approaches fall in two groups, fast semi-numerical methods \citep{Thomas09, Santos10, Mesinger11,Visbal12, Fialkov14} and slower full radiative transfer simulations \citep[e.g][]{Mellema06,Baek10,Zahn11,Semelin17}. The former make a number of approximations while the latter are limited by their resolution given the available computing power. During the Cosmic Dawn, the two processes that shape the 21-cm signal are the Wouthuysen-Field coupling and the heating of the IGM by X-rays \citep[see][for a review]{Furlanetto06}. In this work we will focus on the modelling of the Wouthuysen-Field coupling, also called Lyman-$\alpha$ coupling.

The 21-cm differential brightness temperature is related to the hydrogen spin temperature by: 
\begin{multline}
\delta T_b \,=\, 27 \,\,x_{\mathrm{H_{I}}} \,(1+\delta)  \left( T_s - T_{\mathrm{cmb}}(z) \over T_s \right) \left( 1 + {1 \over H(z)} {d v_{||} \over dr_{||}} \right)^{-1} \\ \times  \,\left( {1+z \over 10} \right)^{1 \over 2}   
\left({\Omega_b \over 0.044} {h \over 0.7} \right) \left({\Omega_m \over 0.27 }\right)^{1 \over 2} \,\,\,\mathrm{mK},
\label{dtb_eq}
\end{multline}
where $x_{\mathrm{H_{I}}}$ is the local neutral fraction of hydrogen, $\delta$ is the overdensity of the gas, $T_s$ is the local spin temperature of hydrogen, $T_{\mathrm{cmb}}$ the CMB temperature, $H(z)$ the Hubble parameter, $d v_{||} \over dr_{||}$ the velocity gradient along the line of sight, $z$ the redshift, and where the usual notation for cosmological parameters is used. The local value of the spin temperature is the result of three competing processes, thermalisation with the CMB, collisions with other particles that drive it to the local kinetic temperature of the gas, and pumping by Lyman-$\alpha$ photons \citep[see][for details]{Furlanetto06} that drives it to the colour temperature of the radiation around the Lyman-$\alpha$ wavelength. The large ($\sim 10^6$) Gun-Peterson optical depth of the IGM during the CD allows the radiation spectrum around the Lyman-$\alpha$ line centre to reach thermodynamical equilibrium with the gas through the many scatterings: thus the colour temperature of the radiation spectrum is almost identical to the gas kinetic temperature. As a result the spin temperature can be written:

\be
T^{-1}_S={T^{-1}_{\mathrm{cmb}}+x_\alpha T_K^{-1} +x_c T_K^{-1} \over 1+x_\alpha+x_c}
\ee
where $x_c$ is the collisional coupling coefficient (negligible at $z<25$) and $x_a$ is the Lyman-$\alpha$ coupling coefficient defined by:

\be
x_\alpha={4 P_\alpha T_{21} \over 27 A_{10} T_{\mathrm{cmb}}}
\ee
and 
\be
P_\alpha=4 \pi \int J_\nu(\nu) \sigma(\nu) d\nu
\ee
where $T_{21}$ is the 21-cm hyperfine transition excitation temperature, $A_{10}$ the corresponding spontaneous emission coefficient, $J_\nu(\nu)$ is the local angle averaged specific intensity and $\sigma(\nu)$ the Lyman-$\alpha$ line cross-section. As we can see, $J_\nu(\nu)$ is the main quantity, along with $T_K$, that can induce fluctuations in $T_S$. Thus modelling the Lyman-$\alpha$ coupling means modelling $J_\nu$.

 Handling the full radiative-transfer equation in an expanding non-homogeneous universe when line scattering and velocity gradients are involved is a daunting task. Even under simplifying assumptions \citep[e.g.][]{Loeb99} the equation is complex. It is easier to get a picture of the involved processes by calculating at the propagation of a single photon. A detailed discussion can be found in \citet{Semelin07}, here we only summarise the main aspects. The physics of the transfer is encapsulated in a single quantity, the optical depth:

\be
\tau=\int_0^L \!\! \int_{-\infty}^\infty \!\!\!\! n_{\mathrm{HI}}(l)\,P(u_{\|},T_{\mathrm{gas}}(l))\,\sigma\left(\nu(l)\times\left[1-{\mathrm{v}_{\|}(l)+u_{\|} \over c}\right]\right) d u_{\|} dl \, ,
\ee
where $n_{\mathrm{HI}}$ is the local neutral hydrogen number density, $\sigma$ is the Lyman-$\alpha$ scattering cross section {\sl in the atom rest frame}, $\nu(l)$ is the redshifting photon frequency in the global rest frame, $\mathrm{v}_{\|}(l)$ is the component of the local gas velocity in the global rest frame parallel to the direction of propagation, $u_{\|}$ is the parallel component of the atom's velocity in the gas local rest frame, and $P$ is the probability for the scattering atom to have a velocity $u_{\|}$ as dictated by the local thermal velocity distribution. There are two typical values that shape the Lyman-$\alpha$ transfer in the high-redshift universe. A photon redshifting from far in the blue wing of the line will accumulate a $\tau \sim 1$ while still $10$ cMpc away from the location where it would redshift into the core of the line (at $z \sim 10$, in a homogeneous universe). Thus, scatterings in the wing of the line introduce a $\sim 10$ cMpc diffusion scale compared to a free streaming case \citep[e.g.][]{Chuzhoy07,Semelin07}. Conversely, when the photon reaches the core of the line, it has a mean free path of less than $1$ ckpc and the optical depth to redshift out of the line is of the order of $10^6$. Thus, wings scatterings determine where a photon will reach the core and core scatterings dominate the overall budget of $P_\alpha$ but are mainly local. This picture applies to a homogeneous expanding universe.

However, we can see how the local value of the gas density, ionisation state, temperature and velocity  all enter the computation of $\tau$. Using local values instead of cosmic averages obviously modifies the computed $\tau$ value. In this work we will quantify the impact on the computed $P_\alpha$ . 

In the semi-numerical approach \citep{Mesinger11, Fialkov14}, $J_\nu(\nu_\alpha)$ at redshift $z$ is computed using a series of Fast Fourier Transforms (FFTs) that implement the convolution of a propagation kernel with the emissivity fields at redshifts $z_{emit} > z$. In the original method, the kernel is built from a free-streaming approximation in a homogeneous medium: photons travel in a straight line until they cosmologically redshift into the line core. The homogeneity assumption makes the kernel spherically symmetric (with a Dirac radial dependence peaked at a radius whose value is determined by $z$, $z_{emit}$ and the cosmology). A recent improvement \citep{Reis21} modifies the shape of the kernel to include the effect of scatterings in wings of the Lyman-$\alpha$ line, although still assuming a homogeneous medium. The SPINTER code presented in this work follows the same approach.

The Lyman-$\alpha$ coupling can also be computed through Monte Carlo radiative transfer simulations \citep[see e.g., using the LICORICE code,][]{Semelin07,Baek09,Vonlanthen11}. In this case $P_\alpha$, the number of scatterings per atom per second is directly evaluated. However, computing an average of $10^6$ scatterings per photon and a sufficient number of photons (to control the sampling noise) reaching the core of the line in each resolution element at each desired output redshift is computationally not yet feasible. As a consequence photons are propagated (through an inhomogeneous medium) until they redshift into the core of the line and then a prescribed number of scatterings is tallied locally, considering that the  spatial diffusion in the core of the line is negligible. \citet{Baek09} give a prescription for this number, based on Monte Carlo simulation in controled environments. 

In this work we show that the prescription by \citet{Baek09} for the number of scatterings in the core  was incomplete and should be modified in the presence of gas velocities. More generally, we will re-examine the impact of the different approximations made in the semi-numerical approach on the resulting $21$-cm signal, using the (corrected) radiative transfer simulation as a proxy for the ground truth. We will focus especially on the effect of gas velocities that seem to have the largest impact. In section 2, using a Monte Carlo modelling including all core scatterings, we quantify the impact of gas velocities in an environment designed on purpose. In section 3, we present the semi-numerical SPINTER code and remind the reader of the main features of the radiative transfer LICORICE code. In section 4, we evaluate the impact of the various approximations in SPINTER in a realistic CD setup. Section 5 presents our conclusions.

\section{The impact of gas bulk-velocity on the Lyman-$\alpha$ coupling}

\subsection{The treatment of gas bulk-velocities in existing methods}
The radiative transfer in the Lyman-$\alpha$ line in a cosmological context has been studied analytically in several works \citep[e.g.][]{Rybicki94,Chen04,Chuzhoy06,Furlanetto06b,Hirata06,Meiksin06}. In most cases, the authors assume a homogeneous and isotropic universe, thus neglecting both the effect of Doppler shifts from the gas bulk velocity and the effect of fluctuations of the density and temperature. Many of these studies focus on the back-reaction from atomic recoil and spin exchanges on the Lyman-$\alpha$ spectrum near the centre of the line, using a Fokker-Planck formalism. These results should still apply if gas bulk velocities are accounted for by modifying the $J_\infty$ (the angle-averaged specific intensity by number of photon "far" from the line centre, or alternatively, in the absence of back-reaction) and using an effective local Hubble flow that includes the divergence of the peculiar velocity field. \citet{Loeb99} study the transfer around a point source in a uniform Hubble flow, thus removing the homogeneity assumption on the emissivity field but not on the medium of propagation. They note that corrections from peculiar velocities may be necessary.

The semi-numerical approach \citep[][]{Furlanetto06d,Santos08,Mesinger11,Fialkov14} considers the actual, non-homogeneous emissivity field from cosmological sources, but still estimates the effects of propagation through a homogeneous and isotropic universe. As such they compute a local $J_\infty$ that does not account for gas bulk velocities. Full Monte-Carlo radiative transfer simulations (\citet{Semelin07,Baek09} and subsequent works) do include Doppler shifts from gas bulk velocities. However, when computing the Lyman-$\alpha$ intensity in a cosmological box, to limit the computational cost, the radiative transfer is halted when photons reach the core of the line and a prescribed number of scatterings is tallied locally. Thus the effect of gas velocities are fully accounted for in the wings of the line (thus affecting where the photon will reach the core), but should also enter through the prescription of the number of scatterings in the core of the line. This last contribution was actually not implemented in LICORICE until now. 

\subsection{Expected magnitude of the effect}
To clarify the issue, we can distinguish two regimes where gas velocities have an impact on the radiative transfer in the Lyman-$\alpha$ line: in the blue wing of the line and in the core of the line. 

\subsubsection*{Effect in the wing}
In the blue wing, where the medium is relatively transparent,  Doppler shifts arising from velocity gradients along the path of the photons will add their own contribution to the cosmological redshifting, modifying the time and location where the photon will eventually reach the core of the line in the local rest frame of the gas. 
The ratio of the Doppler to cosmological frequency shifts is, at least locally, equal to:
\be
r= {d \mathbf{v}_{||} \over dl}\times {1 \over H(t)} \, ,
\ee
where $H(t)$ is the Hubble parameter. A constant ratio $r$ along the radial path of a photon emitted from a point source would cause the radius where the photon reaches the core of the line to shrink (or expand if $r$ is negative) by a factor $1 - r$ (for small $r$ values). Indeed, at a distance $L$ from the source, the accumulated velocity difference would be $H(t) r L$, corresponding to a Doppler shift $\Delta \nu=-\nu {H(t) r L \over c}$ to be added to the cosmological shift $\Delta \nu=-\nu {H(t) L \over c} $. Since the frequency at the centre of the core is constant, the $(1+r)$ factor generated by the gas velocity should be compensated by a $(1-r)$ factor applied to the distance from the source $L$. Photons that would reach the core of the line when crossing a surface element $dS$ at a distance $L$ from the source do so at a distance $L(1-r)$. The corresponding surface element is modified by a factor $(1-r)^2$, changing local flux by $\sim 1 + 2r$ (if $r$ is small). In the same way the local photon number density is modified by $ \sim 1 + 3r $. 

\subsubsection*{Effect in the core}
In the core of the line, where the medium is extremely opaque, the average number of scatterings before the photon is finally shifted to the red wing of the line is determined by the frequency shift between two scatterings (and by the gas density). In a uniform expanding universe the number of scatterings is, on average, equal to the well known Gunn-Peterson optical depth $\tau_{GP}={3\Lambda\lambda_\alpha^3 n_{HI} \over 8 \pi H(t)}$ , where $n_{HI}$ is the number density of neutral hydrogen, $\lambda_\alpha$ is the wavelength at the centre of the line, and $\Lambda$ the natural line width \citep{Gunn65}. As we can see the number of scatterings scales as $H^{-1}$, which has been confirmed by Monte Carlo simulations \citep{Baek09}. This scaling is not directly apparent in the analytical solutions to the Fokker-Planck equation given for example in \citet{Furlanetto06b}: in section \ref{sec_validation} we show that the reason is that the $H^{-1}$ scaling is encapsulated in the value of angle-averaged intensity "far" from the line centre $J_\infty$. This is corroborated by the analytical expressions in \citet{Rybicki94, Chugai80} that exhibit the $H^{-1}$ scaling. The correction from the back-reaction of the gas on the Lyman-$\alpha$ spectrum (that also depends on $H(t)$) comes on top of this primary scaling and can be computed independently. In a non-uniformly expanding medium, one can expect that any expansion/contraction due to gas velocities will locally act in the same way as the Hubble expansion, 
and thus that the number of scatterings per photon will be determined by the, effective, local value of the Gunn-Peterson optical depth: 
\be
\tau^{\mathrm{loc}}_{GP}=\tau_{GP}{H \over H +\mathrm{div}(\mathbf{v})/3}. 
\label{tau_GP}
\ee

This ansatz will be tested in section \ref{sec_validation}. Note that, while $\tau^{\mathrm{loc}}_{GP}$ is the total number of scatterings, more than $99.9 \%$ of these occur within a few thermal widths from the core in frequency, where the mean free path is very short, and thus tallying the total number at the location where the core is reached is a very accurate approximation.

\subsubsection*{Expected magnitude of the velocity gradients}

Let us now estimate the expected relative amplitude of the velocity gradient and Hubble parameter during the Cosmic Dawn. From Newtonian perturbation theory, we know that, in comoving coordinates, the overdensity $\delta$ is related to the comoving velocity by the continuity equation: ${\partial \delta \over \partial t}+\nabla.\mathbf{v}=0$. During the matter dominated era, $\delta$ is proportional to the expansion factor $a$. Thus $H(t)= {\partial \delta \over \partial t} {1 \over \delta}$. We can estimate the typical amplitude of density fluctuations at redshift $10$ at the scale of the simulation resolution $L \sim 1$ cMpc from structure formation theory as $\sigma(L) \sim 0.35$ (where $\sigma(L)$ is the variance of the density field smoothed on scale $L$). Then ${\nabla.\mathrm{v} \over H(t)}\sim 0.35$. We checked that this is indeed the typical value that we find in the LICORICE simulations used in this work. Note that the relative effect on the number of scatterings per photon in the core of the line is $1/3$ of this value, and that, in the wing, $0.35$ can only be an upper limit for the ratio $r$, in particular configurations.

Moreover, we should mention that i) the velocity effect in the wing of the line occurs typically on $10-100$ cMpc scales where $\sigma(L)$ is at most a few percent, ii) the effect in the core of the line, that occurs on small scales indeed, will then be smoothed on the scale of the instrument resolution, typically $10$ cMpc for the SKA, where $\sigma(L) \sim 0.05-0.10$ at those redshifts, iii) $\sigma(L)$ decreases as redshift increases so the effect would be smaller at redshift $20$. Nevertheless, this rough estimate hints at a non-negligible contribution. 

\begin{figure*}
    \centering
    \includegraphics[width=\columnwidth]{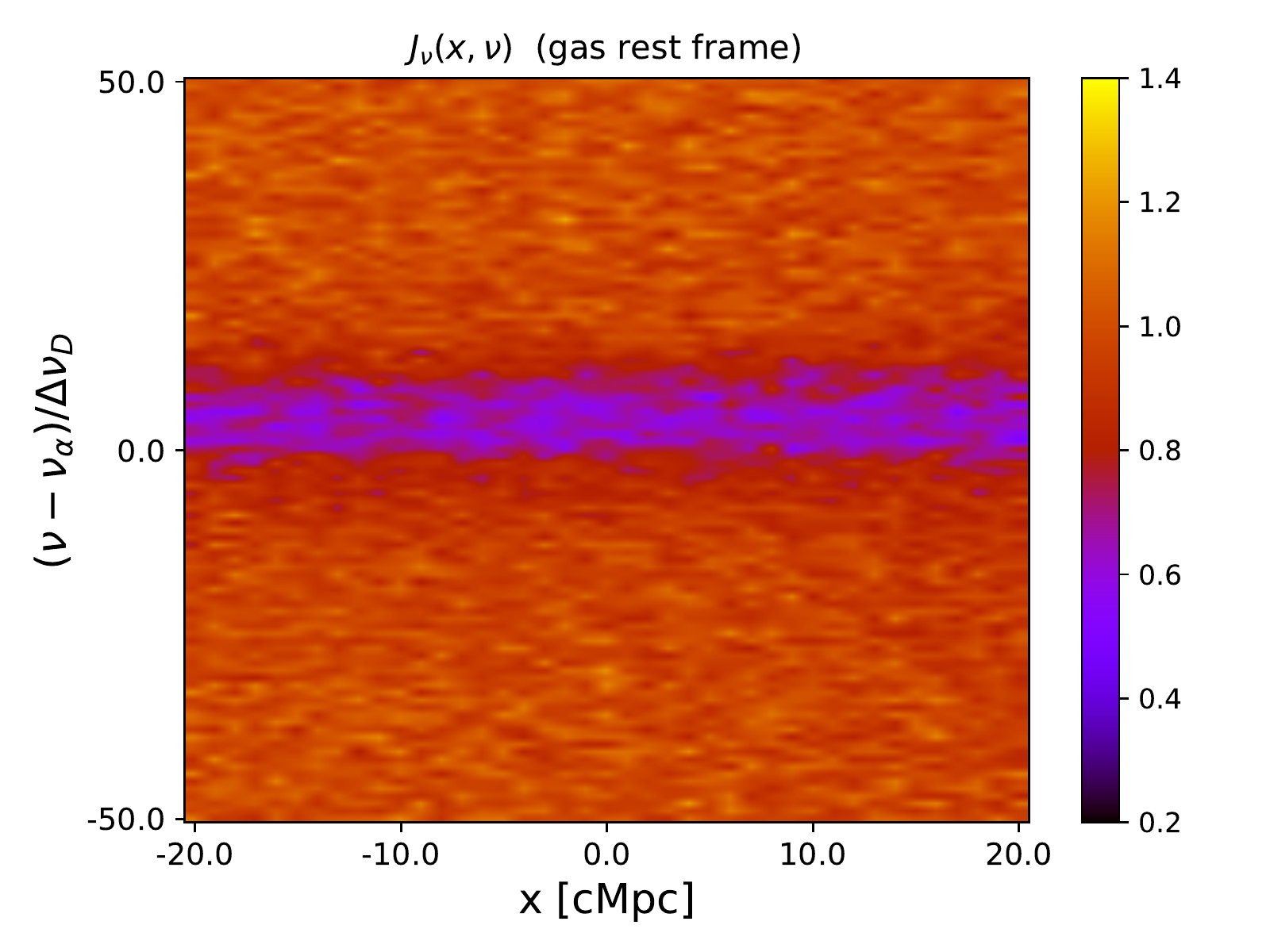}
    \includegraphics[width=\columnwidth]{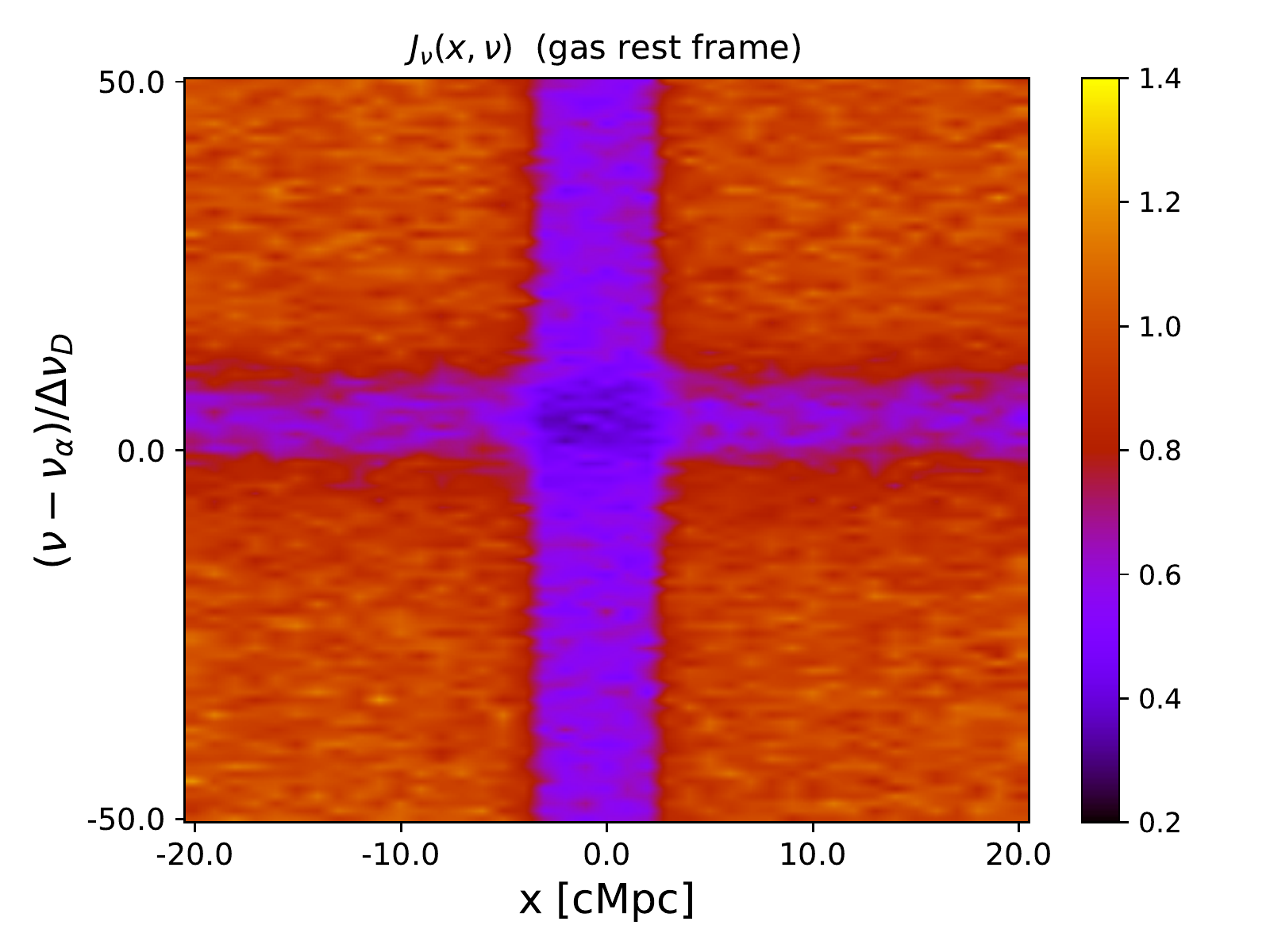}
    \caption{Angle averaged specific intensity near the Lyman-$\alpha$ line centre as a function of the position and frequency in two idealised setups described in the main text as case 1 (left panel) and case 3 (right panel). The angle averaged specific intensity is normalised to $1$ far from the line centre and where there is no effects from velocities. }
    \label{fig:J_test}
\end{figure*}

\subsection{Validating the ${1 \over H +\mathrm{div}(\mathbf{v})/3}$ scaling ansatz }

\label{sec_validation}
As no complete analytical approach exists to describe the radiative transfer in the Lyman-$\alpha$ line in a non-homogeneously expanding medium, we turn to Monte Carlo simulation to validate our ansatz. We use a simplified version of LICORICE, where the spatial dependence of physical fields is prescribed with analytical formulas, thus removing the need for grids and the difficulty of defining a spatial resolution. The Monte Carlo code computes the optical depth along the path of the photons taking into account both the Hubble expansion and the Doppler shifts from the gas velocity field. Thermal motion of atoms are included, scatterings are assumed isotropic in the rest frame of the atom and atomic recoil is computed (but not the back-reaction from spin exchange). More details are given in \citet{Semelin07,Baek09}. To evaluate the effect of the gas velocities gradients on the Wouthuysen-Field coupling we define a geometrically simple situation that nullifies other potential sources of fluctuations.

We consider a cosmological volume with uniform gas density and temperature (computed from a pure adiabatic cosmological evolution). The photons are emitted with a position $(x_{ini},0,0)$ with $x_{ini}$ uniformly sampled between $-130$ and $-70$ cMpc, and an initial direction of propagation along the $x$ axis. The initial frequency is chosen such that the photons will redshift into the Lyman-$\alpha$ line at $z=10$ after having travelled $100$ cMpc. The redshift at emission is sampled in a (narrow) range such that the photon frequency at the desired output redshift ($z=10$ in our case) falls in a range a few hundreds of thermal width around $\nu_\alpha$. To reduce the computing time we ran the test at $1/10$ of the actual gas density (thus reducing the number of scatterings for each photon to $\sim 0.8 \times 10^5$). We have checked that in the absence of velocity fields and atomic recoil, this setup produces a local photon number density at $z=10$ (and thus angle-averaged specific intensity) for any $x$ coordinate between $-20$ and $20$ cMpc that is spectrally flat in a range of $100$ thermal widths centred on $\nu_\alpha$.

Then we consider several possible velocity fields: 
\begin{itemize}
    \item Case 1: no peculiar velocities
    \item Case 2:
    \begin{itemize}
      \item[] if $x \in [-3,3]$ $\,\,\,\,\,\, \mathbf{v}=3yH \,\,\mathbf{u_y} $ 
      \item[] if $x \in [-\infty,-3]$ $\mathbf{v}= 0$
      \item[] if $x \in [3,\infty]$ $\,\,\,\,\,\,\, \mathbf{v}= 0$
    \end{itemize}
    \item Case 3: 
    \begin{itemize}
      \item[] if $x \in [-3,3]$ $\,\,\,\,\,\, \mathbf{v}= yH \,\,\mathbf{u_y} +  z H \,\,\mathbf{u_z} $ 
      \item[] if $x \in [-\infty,-3]$ $\mathbf{v}= 0$
      \item[] if $x \in [3,\infty]$ $\,\,\,\,\,\,\, \mathbf{v}= 0$
    \end{itemize}
    \item Case 4: 
    \begin{itemize}
      \item[] if $x \in [-3,3]$ $\,\,\,\,\,\, \mathbf{v}= (x+3)H \,\,\mathbf{u_x}$
      \item[] if $x \in [-\infty,-3]$ $\mathbf{v}= 0$
      \item[] if $x \in [3,\infty]$ $\,\,\,\,\,\,\, \mathbf{v}= 6H \,\,\mathbf{u_x}$
    \end{itemize}
\end{itemize}

In the above formulas, $H$ is the Hubble parameter at redshift $z = 10$, and the coordinates are in units of comoving Mpc. In cases 2 (resp. 3), the divergence of the velocity field in the slab $x \in [-3.,3]$ equals (resp. equals $2/3$ of) the contribution from the Hubble flow (that is $3H$), and is $0$ elsewhere. In these cases, where the velocities are perpendicular to the initial direction of propagation, there should be little effect from the free streaming regime, but a strong effect from the core scatterings. In case 4, effects from the free streaming regime and from the core scattering should combine. Note that the non-zero uniform velocity at $x > 3$ is chosen only to ensure continuity at $x=3$. Since the gradient is zero in that region, we do not expect any net effect on $J_\nu$. The velocities considered here are obviously larger than expected in a typical cosmological situation (especially being coherent on such large scales), but the goal here is simply to make the effect more visible to validate the ansatz.

\begin{figure}
    \centering
    \includegraphics[width=\columnwidth]{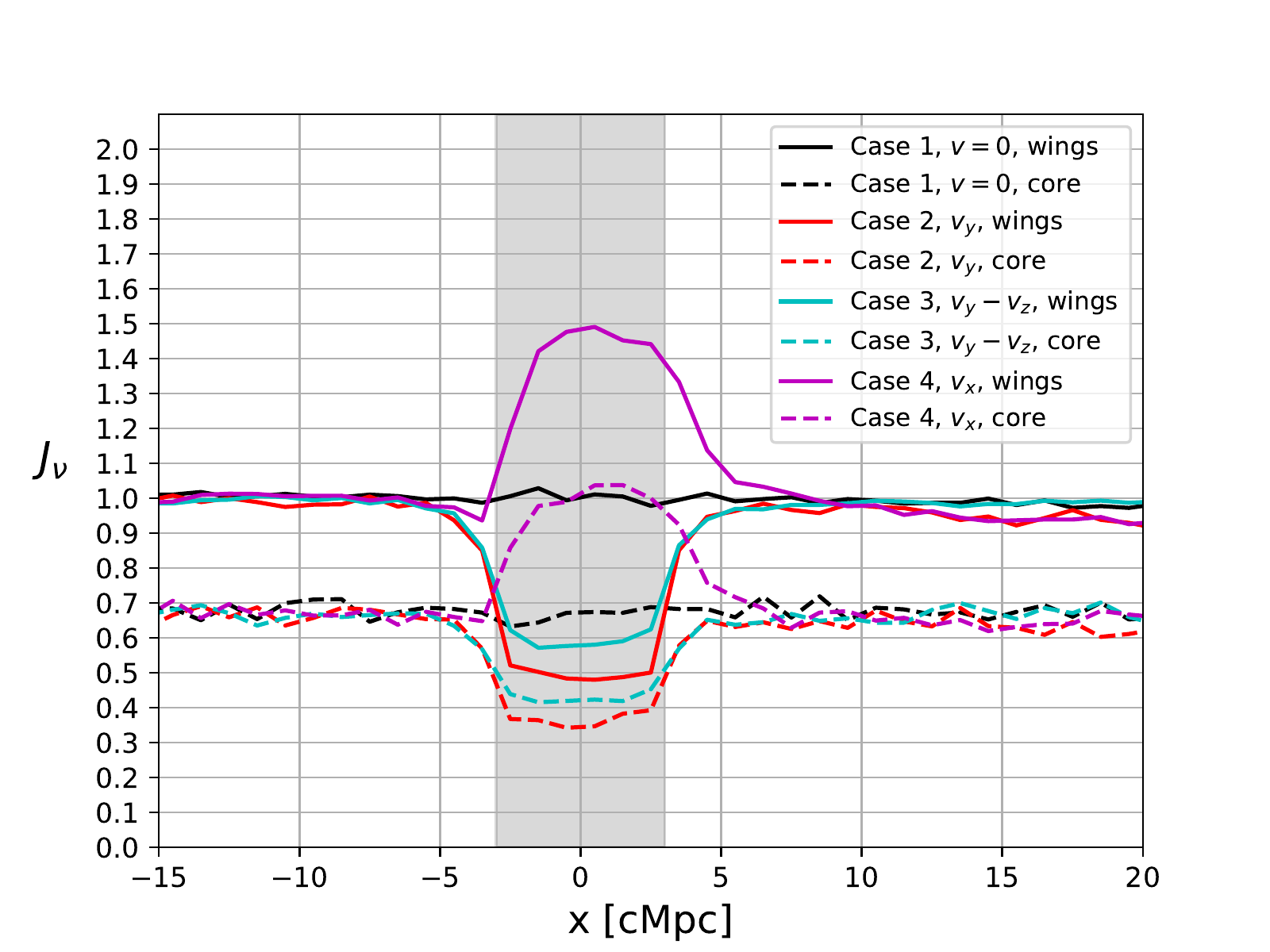}
    \caption{Spatial fluctuations of the angle averaged specific intensity at the centre of the Lyman-$\alpha$ line and in the wings for all 4 cases described in the main text, characterised by different velocity fields in the shaded slab. These correspond to horizontal cuts in fig. \ref{fig:J_test}. }
    \label{fig:avg_J_x}
\end{figure}

In fig. \ref{fig:J_test} we show the normalised, angle-averaged specific intensity $J_\nu(x,\nu)$ at $z=10$ in cases $1$ (no velocities) and $3$ (a slab with velocities perpendicular to the initial direction of propagation). The spectral number density of photons was actually computed, but it differs from $J_\nu(x,\nu)$ only by a factor ${4 \pi \over c}$. On the left panel (no velocities), the effect of the back-reaction from atomic recoil is clearly visible as a depletion of the spectrum around $\nu_\alpha$ (keep in mind that we are operating at one tenth of the cosmological gas density, so the feature is narrower than at the nominal density). On the right, the slab between $-3$ and $3$ cMpc is expanding perpendicularly to the $x$ direction, creating an effective expansion rate $5/3$ as large as in the rest of the universe. We can check that, far from the line centre, $J_\nu(x,\nu)$ is depleted by a factor of $\sim 3/5$ (see fig \ref{fig:avg_J_x} for a more quantitative view), confirming the effect of velocities on the $J_\infty$ of the Fokker-Planck theory. The magnitude of the depletion is consistent with the scaling ansatz, as in case 3, in the centre slab, ${1 \over H +\mathrm{div}(\mathbf{v})/3}={3 \over 5H}$ compared to $1 \over H$ in case 1. The spectrum is further depleted near the centre of the line by the gas back-reaction.
 
\begin{figure}
    \centering
    \includegraphics[width=\columnwidth]{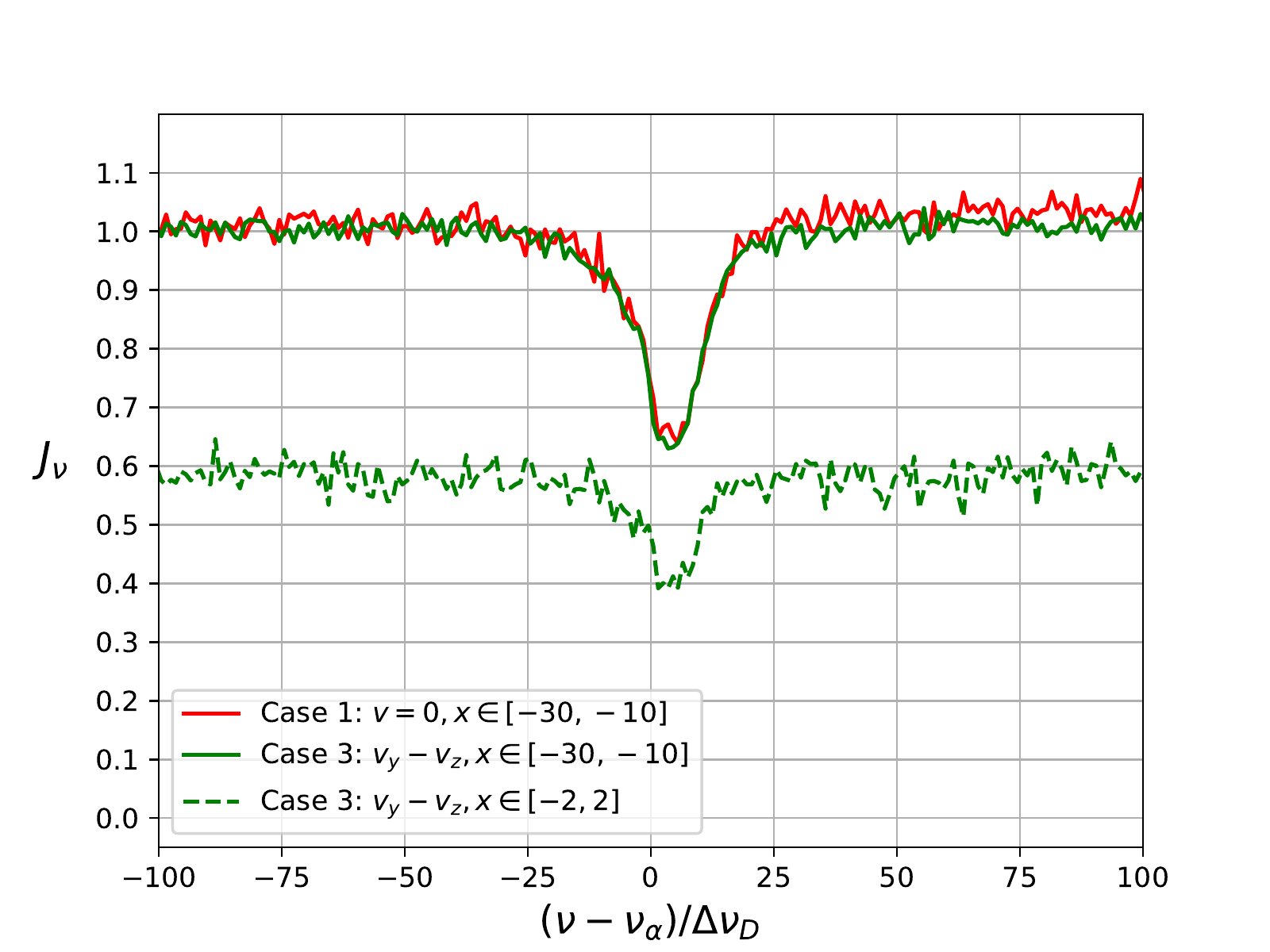}
    \caption{Normalised spectra around the Lyman-$\alpha$ line in different regions of case 1 and 3 (see main text) with either zero velocities (full lines) or an expanding peculiar velocity field (dashed line).These correspond to vertical cuts in fig. \ref{fig:J_test}}
    \label{fig:avg_J_nu}
\end{figure}
 
Fig. \ref{fig:avg_J_nu} shows cuts of the previous maps along the frequency direction, that is normalised spectra. Spectra outside and inside the slab with non-zero velocities are plotted. The $3/5$ scaling resulting from the $5/3$ effective expansion rate is confirmed. Fig. \ref{fig:avg_J_x} shows cuts along the spatial direction, both in the wings of the line and in the core. In the wing cuts, cases 2 and 3 show a $1/2$ and $3/5$ depletion in the non-zero velocity slab, in line with the ansatz and their effective expansion rate. The effect seems to be, as expected, sensitive to $\mathrm{div}(\mathbf{v})$ only and not to the specific topology of the velocity field.
The depletion is similar in the core, showing that the additional impact of velocities on the amplitude of the back-reaction trough the Gunn-Peterson optical depth is small. The not-so-sharp transitions at the boundaries of the velocity slab is due to the scatterings in the wings of the line that create a diffusion in the location where photons reach the core of the line. Case 4 is slightly more complex to interpret. The velocity field, oriented along the free-streaming direction of propagation, induces an effect both in the free-streaming regime and on the core scatterings. The velocity gradient along the (main) direction of propagation in the non-zero velocity slab equals the Hubble parameter. Thus photons that reach the gas-rest-frame core in the [-3,3] slab would otherwise reach the core in a $[-3,9]$ slab. The same is true for reaching any narrow rest-frame range of frequencies. This is just from redshifting and Doppler effects and would, alone, boost $J_\nu$ by a factor of $2$. However, the diffusion regime effect applies another ${3H \over 3H +\mathrm{div}(\mathbf{v})}=3/4 $ correction, leading to the $\times 1.5$ observed correction. This agreement is consistent with both a $1+ {d \mathbf{v}_{||} \over dl}/H$ correction factor in the wing and the $(H +\mathrm{div}(\mathbf{v})/3)^{-1}$ scaling for the number of scatterings in the core.

We will now need to evaluate the impact of the correction due to velocity field in a realistic cosmological case, where the amplitude of the velocities are typically smaller than in this set setup.

\section{Numerical methods}

To run the cosmological test, we will use two different codes: LICORICE, a full radiative transfer code, and SPINTER, a fast code using FFTs that applies a kernel to the emissivity field that takes scatterings into account but assumes homogeneity and isotropy for the gas.

\subsection{ The full radiative transfer with LICORICE}
The LICORICE code performs the full Monte Carlo 3D radiative transfer in the Lyman-$\alpha$ line. It is described in detail in \citet{Semelin07, Baek09, Vonlanthen11} and has been used in a number of subsequent papers to compute the 21-cm signal during the Cosmic Dawn \citep[e.g][]{Semelin17}. We give here a few relevant features of the code and refer the reader to the above references for a complete description. The Lyman-$\alpha$ part of the code performs Monte Carlo ray-tracing on a uniform grid. Directions, frequencies and target optical depths of photons packets are sampled from the relevant distributions. Source luminosities are implemented in such a way that they are not affected by the Monte Carlo sampling noise (i.e. photons are  assigned to sources in a deterministic way, in proportion to their luminosity). Optical depths are computed along the path of the photons, taking into account the local density, ionisation state and temperature of the gas. The effect of the local proper velocity field and of cosmological redshifting on the frequencies of the photon (in the gas rest-frame), and thus on the value of the cross-section for scattering, are taken into account. Indeed, the full Lyman-$\alpha$ line profile, including the wings, is used. Cascades from higher Lyman lines are also included \citep[see][]{Vonlanthen11}. Although this is optional in the code, in typical 21-cm simulations, photon propagation is stopped at the location where they enter the core of the line and a calibrated number of scatterings is assigned to the corresponding cell. This number is the average number of scatterings required to redshift through the line core at the local density \citep[see][]{Baek09}. Indeed, actually propagating all photons until they redshift out of the line would be typically a thousand times more expensive (in proportion to the number of computed scatterings), reaching the $10^7$ CPU hours range for a grid with moderate resolution. The back-reaction from the gas on the local Lyman-$\alpha$ spectrum, due to the atomic recoil and spin-exchange, is computed using the method described in \citet{Hirata06}.

As the LICORICE code results will serve in this work as a proxy for the ground truth it is important to mention the limitations of the code. The most obvious one is the Monte Carlo sampling noise. This is even more the case than in situations where the photon packets can deposit a fraction of their content in each cell along their path such as for ionising photons  or X-rays radiative transfer. For Lyman-$\alpha$ transfer, the scatterings occur essentially in the core of the line where the diffusion length is of the order of $1$ ckpc, and the contribution of each photon  is concentrated in one cell (while wing scatterings are important to determine in which location a redshifting photon will reach the line core, their contribution to the total scattering budget is negligible). Consequently the relative level of the noise scales as $\left({N_\mathrm{phot} \over N_\mathrm{ cell}}\right)^{-{1 \over 2}}$, where $N_\mathrm{phot}$ is the number of photon packets that reach the core of the line in the  time interval over which we want to estimate the average of the Lyman-$\alpha$ coupling, and $N_\mathrm{cell}$ is the number of resolution elements. We see that reaching an average noise level of $1\%$ on a $256^3$ grid already requires $\sim 1.6 \times 10^{11}$ photons. In 21-cm simulations of the Cosmic Dawn, we usually accept larger levels of noise and average the coupling estimation over a time interval of the order of $20$ Myr. The SPINTER code, on the other hand, has no Monte Carlo noise and yield an estimate of the instantaneous coupling. Thus in this work, for the cosmological comparison tests, we will use a $\sim 1.5$ Myr averaging time and we will ensure a 1-2 $\%$ typical noise level\footnote{In practice we create $3.2 \times 10^{12}$ photons, but actually propagate only those whose frequency is such that they will reach the core of the line within the target narrow redshift interval. The reason for this is to minimise the required modifications in LICORICE.}. Reaching that level of noise for a single target redshift on a $256^3$ grid in a cosmological box ($200$ h$^{-1}$ Mpc size) requires around 3000 single-core hours\footnote{on 2015 Intel Xeon E7-8857 CPUs. The code is parallelised with both MPI and OpenMP.}.

\subsection{ The semi-numerical approach with the SPINTER code }

The analytical formula for estimating the average Lyman-$\alpha$ intensity in a homogeneous and isotropic universe and neglecting wing scatterings is described in \citet{Furlanetto06d}. \citet{Santos10} and \citet{Mesinger11} implemented a generalised version that accounts for an non-homogeneous emissivity field. The method to obtain the intensity field at a target redshift comes down to using a series of FFTs to convolve the emissivity field at higher redshifts with a Dirac-like spherical kernel whose radius depends on the emission and target redshifts. \cite{Reis21} improved on that by using a spherically symmetric kernel that accounts for wings scatterings in a homogeneous medium. SPINTER implements a similar approach. In the following sections, we describe a formal framework for using a kernel that includes wings scattering that, to our knowledge, has not been formulated analytically before and give some details on the SPINTER implementation.

\subsubsection{ A Theoretical framework}

In the following, all quantities are comoving unless stated otherwise. Let $\epsilon(\vec{r},t,\nu)\, dV\,d\nu\,dt$ be the number of photons emitted isotropically in a volume $dV$ around position $\vec{r}$, between times $t$ and $t+dt$ and between frequencies $\nu$ and $\nu+d\nu$ (that is $\epsilon$ is the emissivity by number). Let $n$ be the number density of photons such that $n(\vec{r},t,\nu) \,dV \,d\nu$ is the number of photons in a volume $dV$ around position $\vec{r}$ with frequency between $\nu$ and $\nu+d\nu$, at time $t$. We can write the relation between the two quantities resulting from radiative transfer in a very general way as:

\be 
\mathcal{D}n=\epsilon\,.
\ee
Under fairly general conditions, $\mathcal{D}$ is a linear operator (e.g. if two-photon processes are negligible). If a procedure to calculate the Green function of operator $\mathcal{D}$ can be devised, then $n$ can easily be computed for any field $\epsilon$. We will now do so, first in the cosmological free-streaming case and then introducing resonant scattering.

\subsubsection{The free-streaming case}
We will first assume that no absorption or scattering occurs. Instead of the time variable $t$ we will use the redshift $z$. They are related by $dt=-{1 \over H(z)}{dz \over 1+z}$. Let us consider a pulse-like source term $\delta(\vec{r}) \delta(z-z_0) \delta(\nu-\nu_0)$ and the corresponding Green function $G$ for operator $\mathcal{D}$:
\be
\mathcal{D}G(\vec{r},z,\nu)=\delta(\vec{r}) \, \delta(z-z_0) \,\delta(\nu-\nu_0)\,.
\ee
Then from the general theory of Green functions we know that, for any source field $\epsilon$, $n$ can be computed as:
\be
n(\vec{r},z,\nu)=\int \epsilon(\vec{r}_0,z_0,\nu_0) G(\vec{r},\vec{r}_0,z,z_0,\nu,\nu_0) d^3r_0\, d\nu_0\, dz_0 \,.\\
\label{Green_general}
\ee

From the physics of free-streaming radiative transfer in a uniformly expanding universe, we know that photons emitted at redshift $z_0$ and frequency $\nu_0$ will, at time $t$ and redshift $z$, have redshifted to frequency $\nu={1+z \over 1+z_0}\nu_0$, and will be located on a sphere of radius $r_1(z,z_0)=\int_z^{z_0} {c \over H(z^\prime)} dz^\prime$ centred on the emission point. Thus the Green function is necessarily of the form
\be 
G=A\,\delta\left(\|\vec{r}-\vec{r}_0\|-r_1(z,z_0)\,\right)\,\delta\left(\nu-{1+z \over 1+z_0}\nu_0\right)\,.
\label{Green1}
\ee
To establish the expression of $A$, we can use the conservation of the number of photons. More precisely: the number of photons emitted before time $t$ is equal to the total number of photons present in the universe at time $t$ (because we do not have any absorption term). That is:
\be 
\int_V \int_{\nu}\int_{-\infty}^{t}\epsilon(\vec{r},t^\prime,\nu) d^3r\,dt^\prime\,d\nu= \int_V\int_\nu n(\vec{r},t,\nu) d^3r\,d\nu \,.
\ee
Injecting equations \ref{Green_general} and \ref{Green1} we obtain the expression for A and we can write the full expression of the Green function: 
\be 
G={1 \over H(z_0)  (1+z_0)}\,{1 \over 4 \pi r_1^2}\,\delta(\|\vec{r}-\vec{r}_0\|-r_1)\,\delta\left(\nu-{1+z \over 1+z_0}\nu_0\right)\,.
\ee
\medskip
Then,  we can use the Green function to write the number density of photons generated by a source function $\epsilon$. Defining $\vec{r}^\prime=\vec{r}_0-\vec{r}$ and $\vec{n}$ as a vector with norm $1$, and performing the integration on $\nu_0$ and in the radial dimension of $\vec{r^\prime}$:
\be
n(\vec{r},z,\nu)= -\int {1 \over H(z_0)  (1+z)} {1 \over 4 \pi} \epsilon\left(\vec{r}+r_1\vec{n},z_0,{1+z_0 \over 1+z}\nu)  \right) d\Omega dz_0\,.
\ee

The physical, angle-averaged specific intensity $J$ is related to the photon comoving number density by $J=(1+z)^3 {c \over 4 \pi} n$. Then:
\be 
J= (1+z)^2 {c \over (4 \pi)^2}\int_z^{+\infty} \!\!\int_\Omega \epsilon\left(\vec{r}+r_1\vec{n},z_0,{1+z_0 \over 1+z}\nu\right)  {1 \over H(z_0)} d\Omega dz_0\,.
\ee
If the emissivity $\epsilon$ is considered homogeneous at a fixed redshift, we can perform the angular integration and recover the expression often used in analytical models:

\be 
J(z,\nu)= (1+z)^2 {c \over 4 \pi}\int_z^{+\infty} \epsilon\left(z_0,{1+z_0 \over 1+z}\nu\right)  {1 \over H(z_0)} dz_0 \,.
\ee

\begin{table*}[]
    \centering
    \caption{Summary table of the physical processes implemented in the different numerical methods to compute the local Lyman-$\alpha$ flux in the IGM in a cosmological setting.}
    \def\arraystretch{1.5}
    \begin{tabular}{lcccc}
         {\bf Method} & \multicolumn{4}{c}{\bf Implemented physics} \\
         & Wing scatterings & Inhomogeneous $\delta$ and T & Doppler in wings & Doppler in core  \\
        \hline
         SPINTER no-wing or 21CMFAST & no & no& no& no \\
         SPINTER or \citet{Reis21} & yes & no & no & no \\
         LICORICE no velocities& yes & yes& no& no \\
         LICORICE with velocities (wing only)& yes& yes &yes & no \\
         LICORICE with velocities (core only)& yes & yes & no&  yes\\
         LICORICE with velocities &yes &yes&yes&yes \\
    \end{tabular}
    \label{tab:only}
    \vskip 0.5cm
\end{table*}

\subsubsection{Introducing resonant scattering}
The goal here is to perform an approximate computation of the angle averaged specific intensity at the centre of the Lyman-$\alpha$ line, $J_\alpha$, taking into account scatterings in the wings of the line (but not the back-reaction from the interaction with hydrogen atoms in the core of the line, this is handled in post-treatment). We make simplifying assumptions:

\begin{enumerate}
    \item We consider that the frequency of photons is unchanged during scatterings {\sl in the cosmological frame}. This means we ignore the effects of thermal velocities of atoms and proper gas bulk velocities. The validity of these assumptions will be evaluated in section 4. 
    As a result, the frequency part of the Green function remains unchanged, as frequency shifts are only due to cosmological redshifting.
    \item We will consider that the spatial part of the Green function is isotropic. This is true only if the medium around the pulse-source is homogeneous. By extension, it assumes that the effect of gas density fluctuation on $J_\alpha$ is small. This assumption will also be checked.
\end{enumerate}

Then, introducing the function $F$ to isolate the frequency dependence, the Green function takes the form:
\be 
G=F(\|\vec{r}-\vec{r}_0\|,z,z_0)\delta(\nu_\alpha-{1+z \over 1+z_0}\nu_0)
\ee
with the normalising condition: 
\be 
\int F(r,z,z_0) 4 \pi r^2 dr = \left[H(z_0)(1+z_0)\right]^{-1}.
\ee

The main innovation in \citet{Reis21} and in SPINTER is to compute the function $F(r,z,z_0)$ numerically and approximately with a Monte Carlo simulation of wing scatterings for a single isotropic source emitting N photons at redshift $z_0$ in a homogeneous medium. Note that $F$ depends in principle on the gas density, in practice we take it to be the average density of the universe.

Then the photon number density at redshift $z$ can be written:

\be
 n(\vec{r},z,v_\alpha)=\!\!\int \!\! F(\|\vec{r}-\vec{r_0}\|,z,z_0)\, \epsilon\left(\vec{r}_0,z_0,\nu_0 \right)  \left(-{1+z_0 \over 1+z}\right) d^3r_0 dz_0 
\ee
\label{sec_resonant}
with $\nu_0={1+z_0 \over 1+z}\nu_\alpha$.

\subsubsection{Implementation}

The first step in SPINTER is to compute the Green functions. In theory a different Green function should be computed for each Lyman line as upper lines also contribute to $J_\alpha$ through cascades. In practice the cross-section of the lines above Lyman-$\alpha$ is smaller and wing scatterings do not occur often. As a consequence, we use the free-streaming approximation for the upper lines. For the Lyman-$\alpha$ line, we use the resonant scattering formalism. For a given output redshift $z$, the two Green functions are tabulated as function of the radius $r$ and the emission redshift  $z_0$ using a simple Monte Carlo ray-tracing method, subject to the assumption described in section \ref{sec_resonant}. The evaluation is fast (i.e. does not require too many Monte Carlo photons) because we assume that the Green functions have a radial dependence only. The thickness of the radial bins is set by the spatial resolution of the desired outputs. The $z_0$ bins are matched to the radial bins assuming a straight-line propagation. 

Subsequently the photon number density $n$ is computed as a sum over all the redshift bins of convolutions of the emissivity field with a kernel computed using the Green functions. The kernel sums the Green function contributions from all included Lyman lines and also implements the periodic boundary conditions (by summing the contributions of several replica of the sources, and thus several evaluations of the same Green function, if required by the box size and values of the Lyman lines horizons). The convolutions are computed in Fourier space. From $n$, $J_\alpha$ and $x_\alpha$ are computed. The back-reaction can also be computed \citep{Hirata06}.

SPINTER is written in Fortran. Parts of SPINTER are parallelised for shared memory architectures using OpenMP: the initial Monte Carlo computation of the Green functions and the computation of the kernel to use in the convolutions. For the FFTs we use the Intel MKL library. Computing the target field(s) at a single redshift on a $256^3$ grid requires $0.5$ CPU hours (on 2015 Intel Xeon E7-8857 CPUs), and a few minutes using several cores. The difference with the $3000$ hours required for a LICORICE run is striking but bear in mind that the $\sim 1 \%$ noise level used in LICORICE for this comparison would not be called for in many applications.

\begin{figure*}
\includegraphics[width=\hsize]{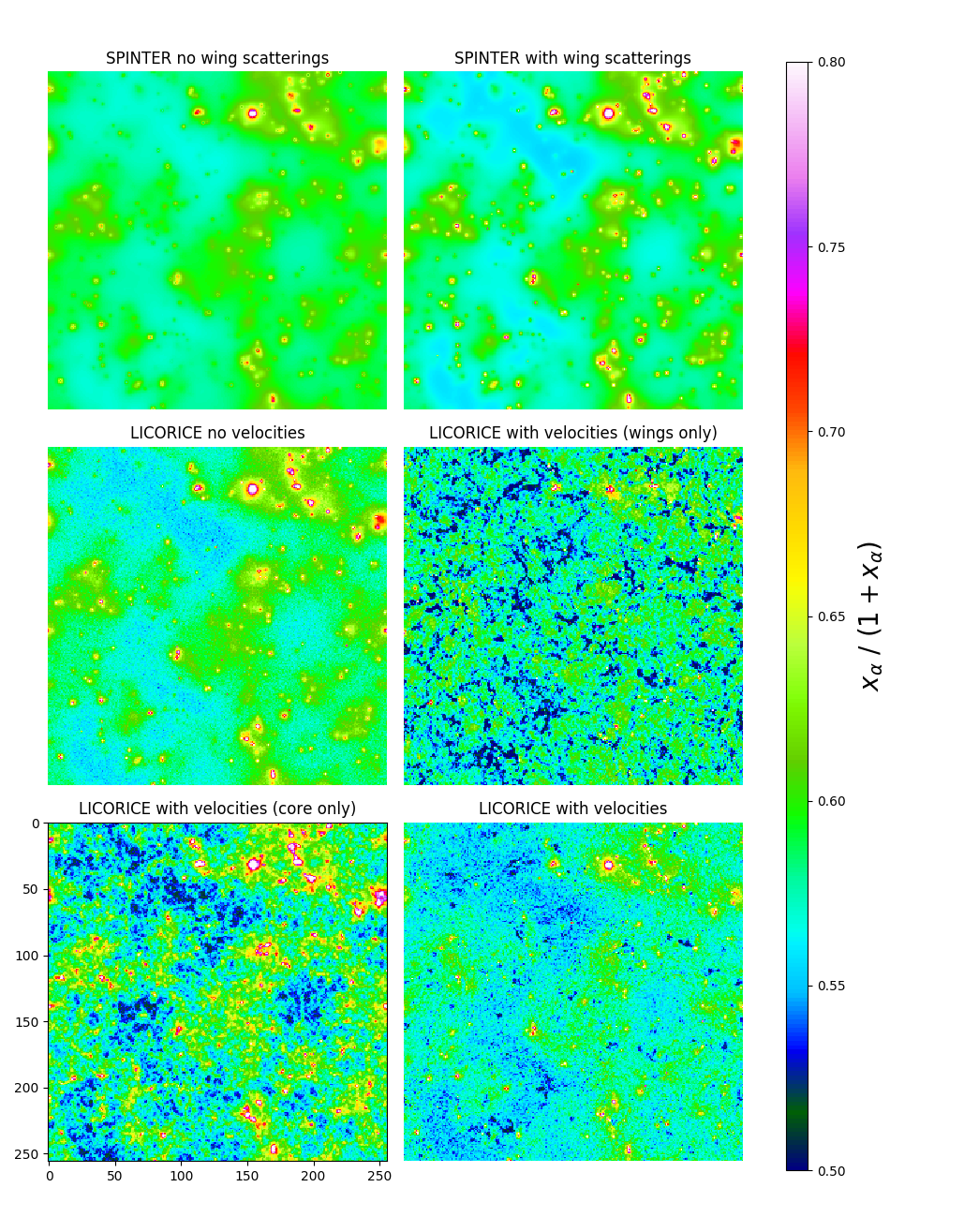}
\caption{Maps of the coupling coefficient x$_\alpha$ (no back-reaction) at $z=12.2$ in our test setup (see main text) for 6 different associations of code and modelling methods. The bottom left panel tick label are in cMpc (the box size is $200$ h$^{-1}$cMpc). }
\label{fig:xa_6maps}
\end{figure*}

\begin{figure*}
\includegraphics[width=\hsize]{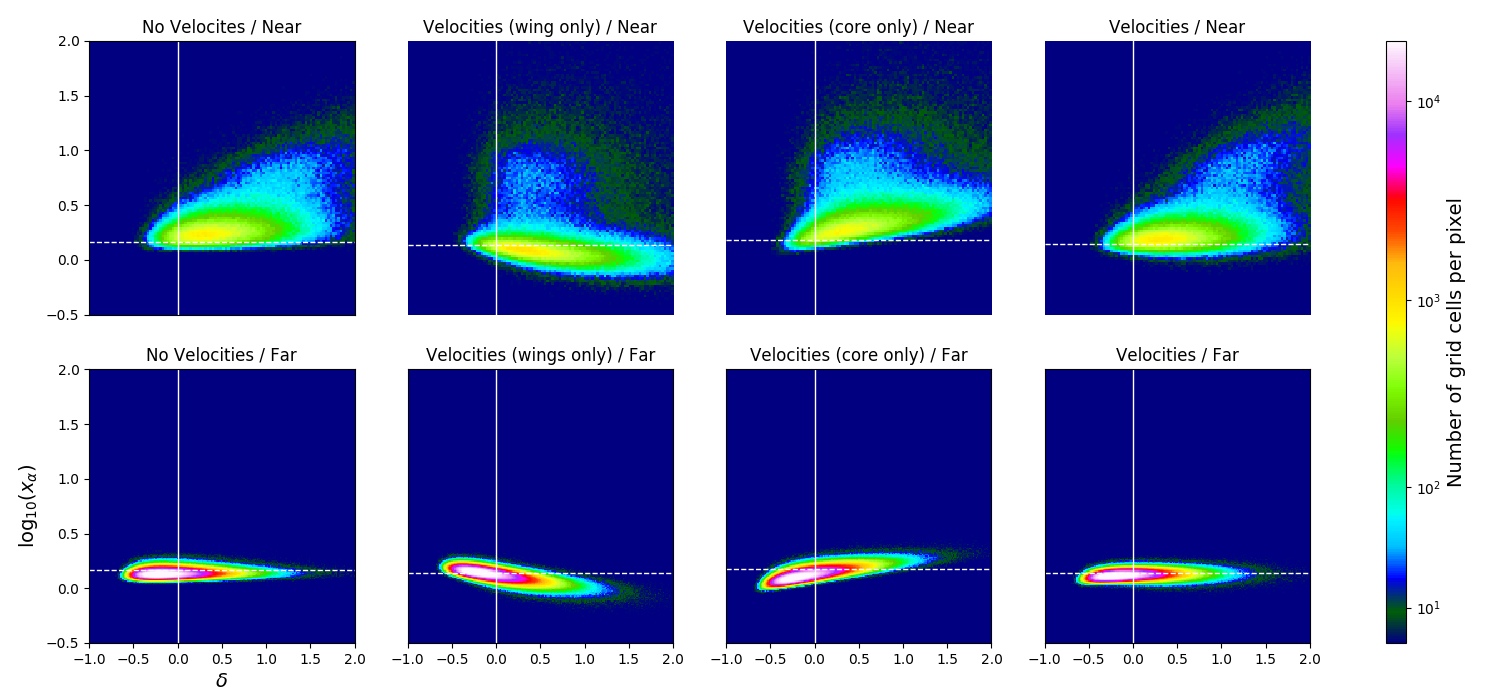}
\caption{ 2D histograms of the number of grid cells as a function of overdensity and $\log_{10}(x_\alpha)$. The upper row is computed including only cells that are located less than $1.2$ cMpc from a source. The low row is computed including only cells that are located more than $6$ cMpc away from any source. The different column corresponds to different modelling methods (the same as in Fig. \ref{fig:xa_6maps}). }
\label{fig:xa_dens_correl}
\end{figure*}

\section{ Evaluating the impact of approximations in SPINTER}
\subsection{Cosmological test setup}
The numerical approaches employed in SPINTER and LICORICE are so different that there are some subtleties involved in comparing their results. Guided by our final goal to be able to robustly model the contribution of the Wouthuysen-Field coupling to the 21-cm signal during the Cosmic Dawn, we choose to compare the three-dimensional field of the coupling coefficient $x_\alpha$ at fixed redshift, before back-reaction (which can be evaluated in post-treatment at a negligible CPU cost) in a cosmological simulation box. The various fields that determine $x_a$ (source emissivity, neutral gas density, temperature and velocity of baryonic matter) are provided by a high resolution radiative hydrodynamics simulation, HIRRAH-21, that resolves halos down to $\sim 4 \times 10^9$ M$_\odot$ in a $200$ h$^{-1}$ cMpc simulation box \citep{Doussot22}. The native resolution of those fields, $2048^3$, are down sampled to $256^3$ (reducing the Monte Carlo noise at $2048^3$ resolution would be prohibitively costly). A difficulty is that SPINTER makes an evaluation of the instantaneous $x_\alpha$ based only on the past history of the emissivity field, while LICORICE, due to the nature of Monte Carlo radiative transfer, evaluates an average of $x_\alpha$ over a redshift interval and takes into account the past evolution of all the other fields as they impacted the propagation of the photons from their emission point to the location where they redshift into the core of the Lyman-$\alpha$ line. To simplify the interpretation of the results we use fields from HIRRAH-21 at an initial redshift $z_{ini}$ and freeze them until the redshift where we want to estimate $x_\alpha$, $z_{final}$. The emissivity is assumed to be zero before $z_{ini}$. We typically choose $z_{final}$ such that photons emitted just below Lyman-$\beta$ at $z_{ini}$ have enough time to redshift down to  Lyman-$\alpha$. Moreover, in the case of LICORICE, we select for actual propagation only photons that will reach a Lyman line in the narrow $\delta z = \pm 0.02$ range around $z_{final}$. Thus we need to sample only a narrow frequency range in the spectrum of the sources, whose boundary change with the emitting redshift. In practice, photons emitted with a frequency around $\nu_{ini}={1+z_{ini}\over 1+z_{final}}\nu_\alpha$  in a range $\delta \nu={\delta z\over 1+z_{final}} \nu_{ini}$  will not necessarily reach the local rest frame $\nu_{\alpha}$ in the target redshift range due to Doppler shifts from the local gas velocities. We still propagate them until they reach $\nu_\alpha$ and count them, considering that they replace photons that would have reached the local $\nu_{\alpha}$ in the target redshift range by having been emitted with a frequency slightly outside of the initial frequency range (we use a flat source spectrum). We are then able to bring the Monte Carlo noise level to just a few percent, even though the output is averaged only over a $\delta z= \pm 0.02$ interval.

We will be using $z_{ini}=12.2$ which is a rather low value to study the Cosmic Dawn. Indeed, at this redshift the volume-averaged $x_\alpha$ is $\sim 1.4$ in the HIRRAH-21 simulation. The reason for this late coupling is the rather high value of the minimum resolved halo mass $4. \times 10^9$ M$_\odot$. HIRRAH-21 is a full radiative-hydrodynamics simulation and reaching this mass resolution in a 200 h$^{-1}$cMpc box is already a challenge. At the same time $x_\alpha \sim 1$ is the regime where fluctuations in the coupling are most likely to dominate the brightness temperature fluctuations. We believe that  the effects we exhibit would be at least qualitatively similar for models where the studied regime occurs at higher redshift.

\subsection{Comparing SPINTER and LICORICE Lyman-$\alpha$ coupling maps}
Let us assume that the signal is seen in absorption at a redshift where the kinetic temperature of the neutral gas, $T_K$, is substantially smaller than the CMB temperature, $x_\alpha$ is of the order of $\sim 1$ and the collisional coupling is negligible. Then we can simplify the computation of the spin temperature to $T_S^{-1}\sim{{x_\alpha}T_K^{-1} \over 1 + x_\alpha}$. Since we then also have $\delta T_b \propto T_S^{-1}$, we find that $ \delta T_b \propto {{x_\alpha} \over 1 + x_\alpha}$. Thus studying $ {{x_\alpha} \over 1 + x_\alpha}$ will give us a good first idea of the impact of the different ways of modelling $x_\alpha$ on the 21-cm brightness temperature.
In fig. \ref{fig:xa_6maps} we show $x_\alpha \over (1+x_\alpha)$ maps (slices with single cell thickness) for $z_{ini}=12.2$, for various modelling choices. 

The first (expected) result is that the map produced with SPINTER including the effects of wing scatterings is more contrasted than the map for SPINTER when the wing scatterings are ignored. Indeed, it has been shown before that the wing back-scatterings create a steeper radial profile of the Lyman-$\alpha$ flux around a point source \citep{Chuzhoy07, Semelin07}. This larger contrast is also found in \citet{Reis21}. The second result is that there is very little difference, apart from Monte Carlo sampling noise, between the map produced with SPINTER with wing scatterings and the one produced with LICORICE when the fluctuations of the HI number density and temperature, but not of the velocity, are included. This seems to indicate that using homogeneous density and temperature fields in SPINTER is an acceptable approximation. We will verify this using the power spectrum. 

However, we do see substantial changes in the maps when the effect of the gas velocities on the propagation of Lyman-$\alpha$ photons is included. To better understand those changes we separate two contributions for the effect of velocities: the effect in the wings that changes the location where the photons reach the core (that depends on the gradient of the velocity along the direction of propagation) and the effect in the core that changes the number of scatterings for each photon before redshifting out of the line (that depends on $\mathrm{div}(\mathbf{v})$). The main effect in the wings is to weaken the coupling close to the sources: there, a large fraction of the local photons comes from the neighbouring source, and since the gas velocity field is converging toward the source the Doppler effect will create a blueshift that will counterbalance the cosmological redshifting. Thus photons will have to travel farther from the source to reach the gas rest-frame Lyman-$\alpha$ frequency. The second visible effect of velocity in the wings is to create rather small scale fluctuations in the voids, where the coupling is the weakest. In the core of the line and near the sources, velocities have the opposite effect as the one they have in the wings: the negative velocity divergence increases the number of scattering and thus the coupling intensity. The effect in the core also creates fluctuations in the voids. The visible consequences of the total velocity effect (wings and core) is i) an overall small decrease of the average coupling (a few percent) ii) a weakened coupling close to the sources iii) small scale fluctuations in the voids. To gain a better grasp on the fluctuations in the voids, we now analyse the correlation to the density field.

\subsection{Correlation of the velocity-induced fluctuations of $x_\alpha$ to the density field}

We showed in section 2 that we expect the velocity gradients (along the line of propagation or acting trough the divergence) to have an impact on the Lyman-$\alpha$ coupling. In the linear regime, the growth of the density field is proportional to the divergence of the velocity. So, we can expect that the fluctuations in $x_\alpha$ caused by gas velocities will correlate with the density field, at least in some regions. We explore this possible correlation in fig. \ref{fig:xa_dens_correl} where we plot 2D histograms of the number of grid cells in a given bin of $x_\alpha$ and overdensity $\delta$, with different contributions from the velocities. The histograms are shown separately for regions more than $6$ cMpc away from any source and for regions closer than $1.2$ cMpc to a source. This split is more relevant than a split based on the density: as we see in the plots, overdense and underdense regions are found both near and far from the sources. 

The modification to $x_\alpha$ caused by including the velocity effect on core scatterings is completely determined by the local value of the divergence of the velocity (see eq. \ref{tau_GP}). Overdense (underdense) regions typically have negative (positive) velocity divergence that should result in increased (decreased) $x_\alpha$. This is exactly what we observe in the third column of fig. \ref{fig:xa_dens_correl}: an increased positive correlation between $x_\alpha$ and $\delta$ is observed compared to the case without velocities (first column). This positive correlation exists both near and far from the sources. 

In the case where the effect of velocities  is included only for the propagation in the wings, the effect depends on the gradient of the velocity along the line of sight. Thus the local effect is direction dependent. Near the sources, we can expect a dominant contribution from photons travelling radially from the closest source and thus, as stated in section 2, a net decrease of $x_\alpha$, compared to a case without velocities. Typical spherical collapse predicts increasing velocity gradients toward the higher-density centre, so in our case a stronger decrease of $x_\alpha$. We do observe this anti-correlation between $x_\alpha$ and $\delta$ in fig. \ref{fig:xa_dens_correl}. What is less anticipated is that the anti-correlation persists far from the sources, where the radiation field is more isotropic (in the case of a perfectly isotropic radiation field, we would expect a zero net effect as photons travelling in opposite directions would experience opposite effects from the velocity field). This seems to indicate that a correlation between the velocity field (and thus the density structures, i.e. pancakes, filaments) and the main direction of propagation of photon is still effective far from the sources. The last column of fig. \ref{fig:xa_dens_correl} indicates that the correlation and anti-correlation induced by the effect of velocities on core and wings transfer do cancel out to a large extent when the two effects are combined.

\subsection{ Impact on the power spectrum of the Lyman-$\alpha$ coupling}
\begin{figure}
    \centering
    \includegraphics[width=\columnwidth]{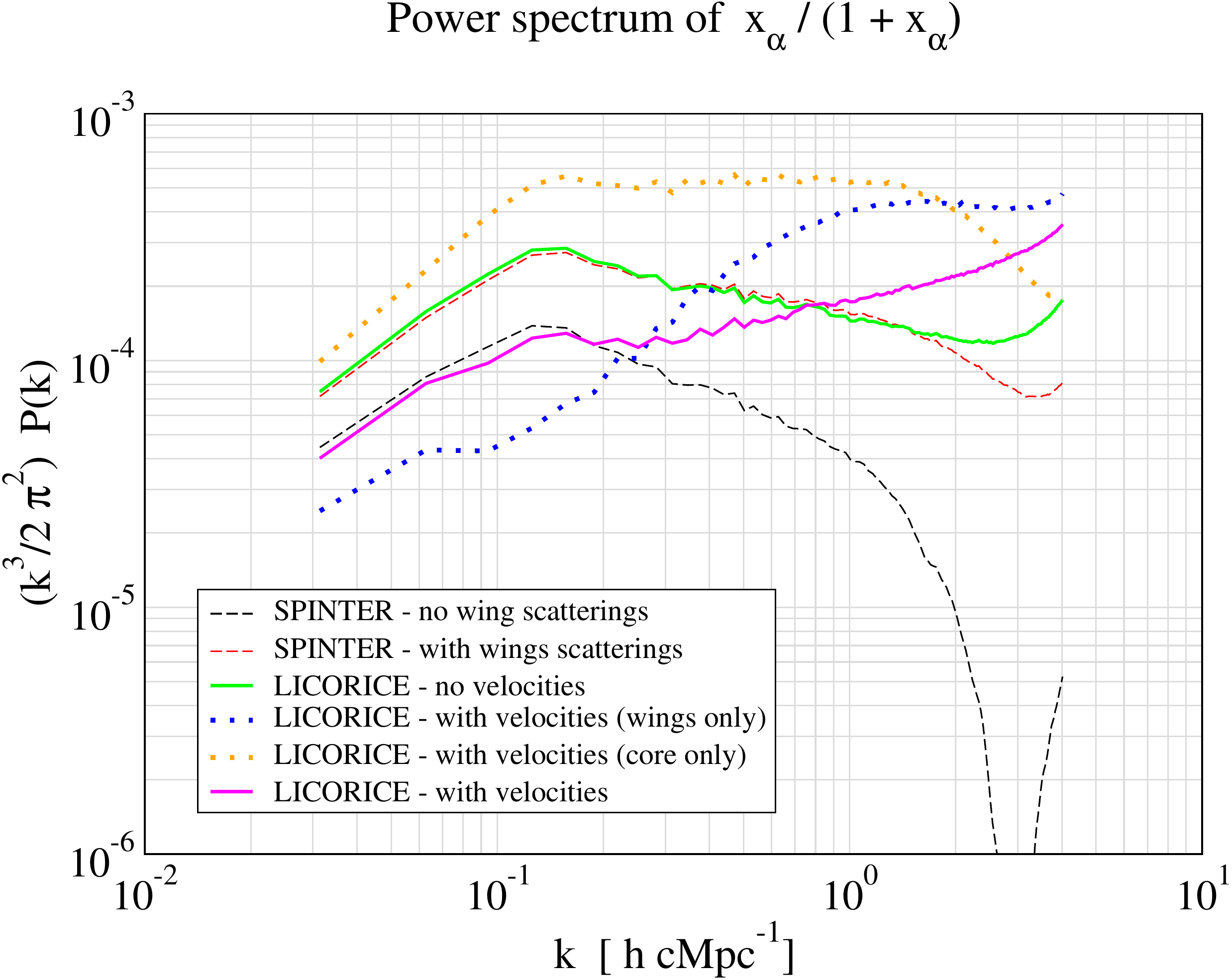}
    \caption{Three-dimensional isotropic power spectra of $x_\alpha \over 1+ x_\alpha$ in a realistic cosmological setting at $z=12.2$ for the different code - modelling method combinations.}
    \label{fig:xa_pwspt}
\end{figure}

Fig. \ref{fig:xa_pwspt} shows the 3D isotropic power spectrum of $x_\alpha \over 1+ x_\alpha$ in the same 6 cases as fig. \ref{fig:xa_6maps}. The same effects that we identified on the maps are present in the power spectra. Quantitatively, including wing scatterings in SPINTER boosts the power on all scales by a factor of 2 or more. LICORICE without gas velocities gives results nearly identical to SPINTER with wing scatterings, except for a small boost at small scales either due to residual Monte Carlo noise or some limited self-shielding effects in high density regions \citep{Semelin07}. Including the effect of gas velocities only for  the number of scatterings in the core boosts the power on all scales (from a factor  $1.5$  on large scales up to a factor $3$ on small scales), while only including the effect of gas velocities on the propagation in the wings decreases the power on large scales (by a factor $\sim 3$) and increases it on small scales (by a similar factor). The total effect of velocities, combining the two contributions, is to decrease the power on large scales and increase it on scales corresponding to wavenumbers larger than 1 h cMpc$^{-1}$. Note that the magnitude of the variations between the different modelling may depend on the history and morphology of the Lyman-band emissivity field (as we check by looking at the same quantities at different redshifts). What matters is that these variations exist and cannot be ignored at least in some specific regimes.

\begin{figure*}
    \centering
    \includegraphics[width=\columnwidth]{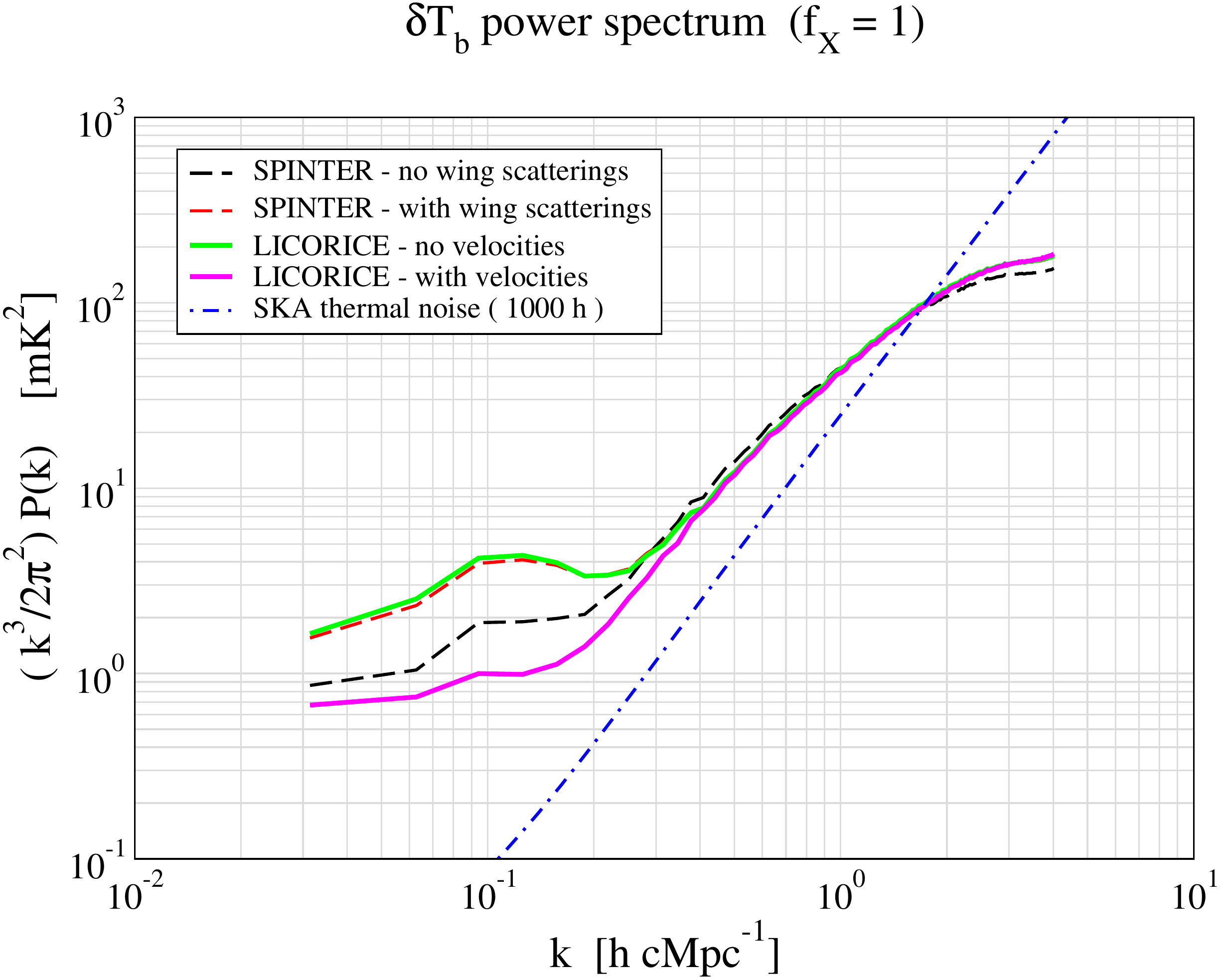}
    \includegraphics[width=\columnwidth]{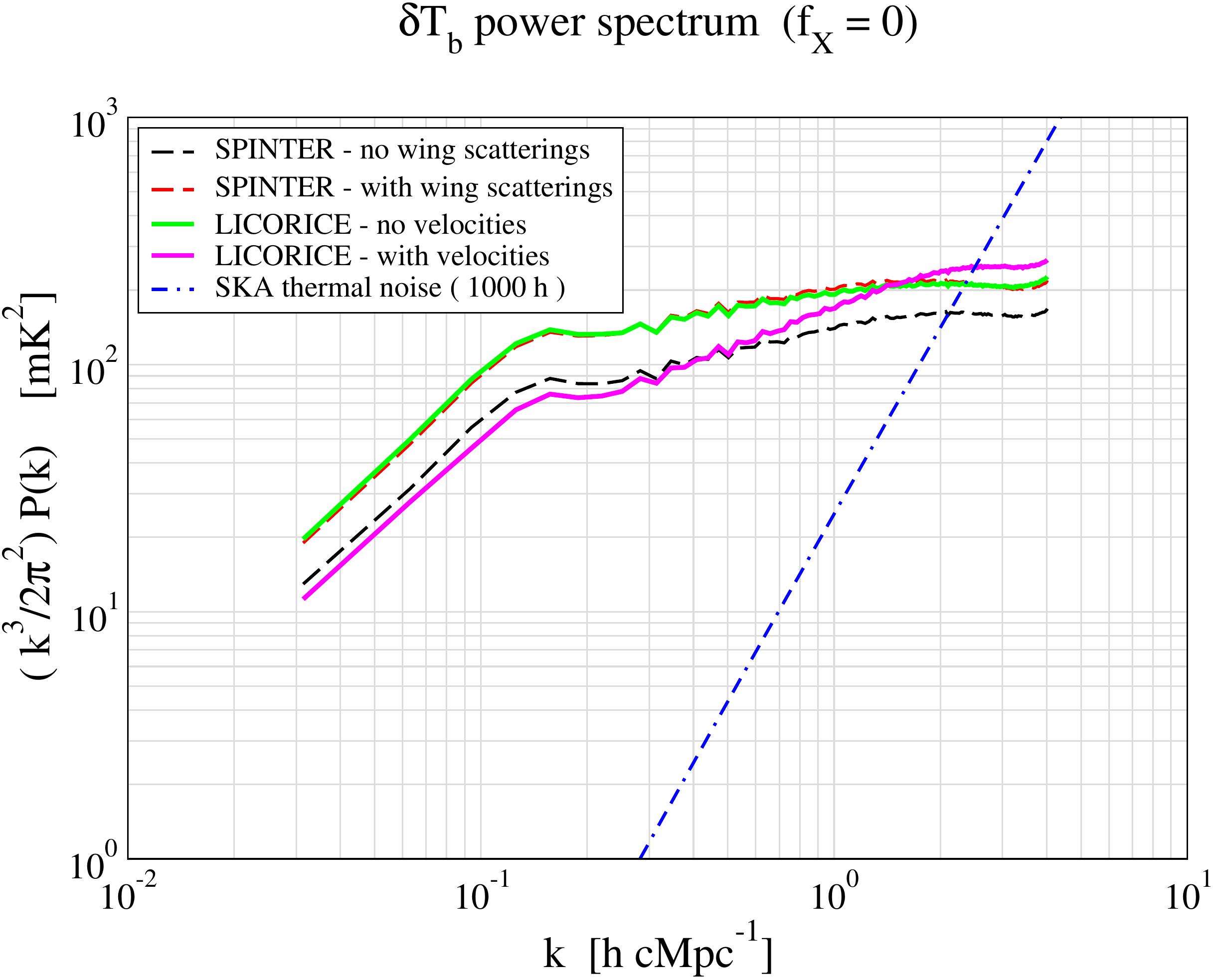}
    \caption{The 3D isotropic 21-cm signal power spectrum for a  realistic cosmological setting at $z=12.2$ (see main text) is plotted for different modelling methods and codes. On the left panel, the IGM has been heated with a fiducial amount of X-rays ($f_X=1$), on the right panel the IGM has been subject to adiabatic cooling/heating only ($f_X=0$). The expected SKA thermal noise level for 1000 hours of observation (see further detail in the main text) is also plotted to quantify how the Lyman-$\alpha$ modelling could bias parameter inference performed on SKA observations.}
    \label{fig:powerspectrum}
\end{figure*}

\begin{figure}
    \centering
    \includegraphics[width=\columnwidth]{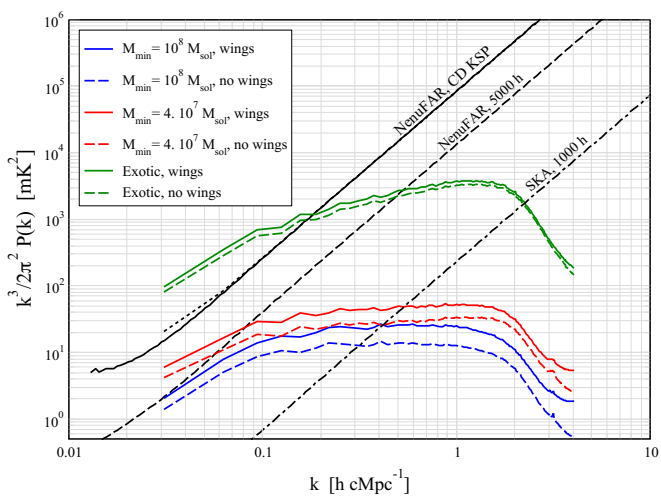}
    \caption{The 3D isotropic power spectra of the 21-cm signal at $z=16.5$ for three different models (see main text for further details) are plotted for two different ways of modelling the Lyman-$\alpha$ coupling (both without the contribution of gas velocities) and compared to the expected thermal noise of observations on the NenuFAR and SKA radio interferometers. The dotted black line shows the contribution from sample variance in the case of the exotic signal.}
    \label{fig:nenuFAR}
\end{figure}

\subsection{Impact on the 21-cm brightness temperature power spectrum}

While the previous study of the impact of gas velocities on $x_\alpha$ is interesting because this quantity is close enough to the physics of radiative transfer that we can readily interpret the results, it is not sufficient to estimate whether the full modelling of gas velocities would change how we infer astrophysical knowledge from $21$-cm power spectrum observations during the Cosmic Dawn. Indeed, in addition to $x_\alpha$ fluctuations, neutral hydrogen density fluctuations and gas kinetic temperature fluctuations determine the brightness temperature power spectrum. Moreover, these fluctuations are clearly not uncorrelated and have varying relative contributions depending on the redshift and on the astrophysical model, as parameterised for example by $f_X$ that quantifies the intensity of heating by X-rays \citep[see e.g.][for a full définition]{Semelin17}. 

In fig. \ref{fig:powerspectrum} we present the 3D isotropic power spectrum of $\delta T_b$ computed from the HIRRAH-21 fields at $z_{ini}=12.2$ for $f_X=1$ and $f_X=0$. The X-ray contribution is a mix of sources with a soft spectrum (like active galactic nuclei) and a hard spectrum (like X-ray binaries) in equal contribution \citep[a parameter $r_{H/S}=0.5$ as defined in][]{Semelin17}. The HIRRAH-21 simulation was run with $f_X=1$. The brightness temperature for $f_X=0$ can be evaluated in post-treatment by recomputing the local kinetic temperature of the neutral gas assuming an adiabatic evolution from a homogeneous universe at $z \sim 130$, the redshift of thermal decoupling between the gas and the CMB. On large scales we observe an effect similar to what we found for the $x_\alpha \over 1+ x_\alpha$ power spectrum: a boost when including wing scatterings in SPINTER and a depletion when including velocities in LICORICE. What happens on smaller scale seems to depend on the presence of X-ray heating. When no X-ray heating is present ($f_X=0$) the $\delta T_b$ power spectrum shows similar reactions to the different approximations as the Lyman-$\alpha$ coupling power spectrum. When X-ray heating is present ($f_X=1$), the $\delta T_b$ power spectrum shows little sensitivity on small scales to the $x_\alpha$ modelling. 

In the simulation with $f_X=1$, the neutral IGM has been heated to an average temperature of $\sim 7$ K, up $2$ K from the $\sim 5$ K in the $f_X=0$ case. Due to the presence of soft X-ray with limited mean-free-path in the neutral IGM, we can expect the heating close to the sources to be even larger, locally initiating the heating transition toward a signal in emission. Then, the power spectrum of the $21$-cm signal may be dominated on small scales by the contribution of these heated bubbles, and thus insensitive, on these scales, to the Lyman-$\alpha$ fluctuations. It is likely that the scale below which the heating fluctuations become dominant depends on the relative contribution of hard and soft X-rays as, typically, a harder spectrum results in heating on larger scales. 

In fig. \ref{fig:powerspectrum}, we also plot, following the methodology of \citet{McQuinn06}, the expected thermal noise level for 1000h of observation with SKA at z=12 (10 MHz bandwidth, width of $k$-bins equal to the $k$ value at the centre of the bin, $T_{\mathrm{sys}}=100 + 300 \left(\, \nu\, /\,150\,\mathrm{MHz}\,\right)^{-2.55}\,\,$K and  $30$-$\lambda$ low $k$ cutoff). Sample variance is not included in the plot. For reference, in this case, it equals $10\%$ of the signal power spectrum at $k=0.03$ h cMpc$^{-1}$ and already decreases to $1.6\%$ at $k=0.1$ h cMpc$^{-1}$. As we can see, SKA should be able to easily discriminate between the various modelling methods. This means that the inference of astrophysical parameters from SKA observations will be biased if an approximate Lyman-$\alpha$ coupling modelling is used. 

\subsection{Observability with NenuFAR and SKA}
The HIRRAH-21 simulation, despite its high resolution for a fully coupled radiative transfer simulation, does not resolve halos with masses less than $\sim 4\times 10^9$ M$_\odot$, whereas we know that halos with masses down to $10^8$ M$_\odot$ should efficiently form stars from hydrogen atomic cooling. This results in a rather late onset of the Cosmic Dawn and of reionization. A complementary assessment of the potential impact of Lyman-$\alpha$ coupling modelling should focus on higher redshift scenarios, to determine how Lyman-$\alpha$ modelling will affect the interpretation of observations. We present such an assessment here making the assumption that the amplitude of the effects of including wing scatterings versus including velocities remains of the same order at higher redshift. Indeed we will show only the effect of wings scatterings because running a full radiative transfer of the Lyman-$\alpha$ with an averaging of the coupling over a sufficiently narrow redshift interval is very difficult while considering a full cosmological evolution (remember that in the previous cosmological test, the fields were frozen at $z=12.2$). If we were to use the redshift interval that we typically use in LICORICE simulations for averaging the coupling, we would not be able to easily distinguish between the effect of velocities and the effect of averaging over a large interval.

The models plotted in fig. \ref{fig:nenuFAR} are computed with a modified version of LICORICE that takes into account halos below the simulation resolution using the conditional mass function formalism in a way similar to 21cmFAST \citep[see][for an alternative approach to including unresolved star formation]{Gillet21}. Resolved halos are treated as before. An unresolved collapsed fraction (including the effect of Poisson noise on the conditional mass function) is computed for each particle and contributes to the star formation. Full details about the implementation will be given in Meriot \& Semelin, in prep. The plotted models use a $200$ h$^{-1}$cMpc box and $256^3$ particles. While they only resolve halos with masses larger than $2. \times 10^{12}$ M$_\odot$, the conditional mass function treatment allows for star formation in unresolved halos with masses down to either $4\times 10^7$ M$_\odot$ or $10^8$ M$_\odot$. A weak X-ray contribution was used, $f_X=0.1$, to maximise the absorption signal. Note however that using $f_X=1$ would not decrease the signal much as it is still insufficient to start the heating transition at the redshift of interest ($z=16.5$).
The "exotic" model includes an additional homogeneous radio background following \citet{Fialkov19}, with an intensity parameter $A_r=9.6$ and spectral index $\beta=-2.6$ (resulting in an additional background $10$ times stronger than the CMB at the redshift of interest), such that the corresponding global signal has an amplitude corresponding to the claimed EDGES detection by \citet{Bowman18} \citep[however see also][]{Singh22}. All models use an $f_\alpha=2$ \citep[see][for a definition]{Semelin17}, a phenomenological parameter that allows to vary the average intensity of the Lyman-$\alpha$ coupling. The power spectra are plotted at $z=16.5$. 

The redshift was selected as the lowest possible for an observation with the NenuFAR radio interferometer while averaging over a $10$ MHz bandwidth. Indeed, NenuFAR\footnote{https://nenufar.obs-nancay.fr/en/homepage-en/} \citep{Zarka12b, Zarka20} has a 10-85 MHz operating bandwidth. In 2019, the Cosmic Dawn Key Science Program\footnote{https://vm-weblerma.obspm.fr/nenufar-cosmic-dawn/} (ES01, P.I. L. Koopmans, B. Semelin, F. Mertens), hereafter CD KSP, started on NenuFAR. It will have accumulated $> 1300$ hours of single field observation by the end of 2022 \citep{Mertens21}. The expected thermal noise of NenuFAR is plotted together with the models. NenuFAR being still under deployment, the CD KSP observations have been acquired with varying (expanding) configurations: $475$ hours with a 56-stations core configuration, $460$ hours with a 80-stations  core configuration, and we expect $\sim 140$ more hours with the 80-stations core configuration and $150$ hours with the full 96-stations core configuration by the end of 2022.  The CD KSP noise level plotted in the figure reflects this mix. It uses the same $\Delta k=k$ bin width and 10 MHz bandwidth as for the SKA, but a $10$-$\lambda$ low $k$ cutoff (due to the different station configuration). It does not include contributions from the $k_{\|}=0$ modes, following \citet{Mertens20}. The system temperature, $T_\mathrm{sys}=T_\mathrm{inst}+T_\mathrm{sky}$, uses the same $T_\mathrm{sky}$ as for SKA but $T_\mathrm{inst}=874$ K, the value given at $80$ MHz by the Nenupy software  \citep{Loh20}. The NenuFAR - 5000 h noise level assumes $5000$ hours of observation on the full 96-stations core configuration, and the SKA $1000$ h uses the same parameters as for fig. \ref{fig:powerspectrum} but at redshift $z=16.5$.

The exotic signal is detectable by NenuFAR CD KSP at wavenumbers $k < 0.1$ h cMpc$^{-1}$, at a level likely sufficient to discriminate between the two Lyman-$\alpha$ coupling modellings. To quantify the discrimination power, one would need to compare the power spectrum of the difference of the models to the thermal noise and sample variance. This would however not be enough to yield an estimate on the induced bias on the model parameters, which is the final issue. We do not, at this time, have a full framework to do this computation. The two more standard signals are only marginally detectable with NenuFAR in $5000$ h at the largest scales and the robustness of this detection would be sensitive to the Lyman-$\alpha$ modelling. Note that the plotted "non-exotic" models are still rather well suited for a detection (high $f_\alpha$, low minimal halo mass for efficient star formation). They are however incompatible with the \citet{Bowman18} claimed detection, since they have a sky-averaged absorption signal of only $\sim 40$ mK at this redshift. Thus NenuFAR can in principle put constraints on the more optimistic models in term of signal strength, but an accurate determination of which models can be excluded and at what level will have to rely on an accurate Lyman-$\alpha$ coupling modelling.

Fig. \ref{fig:nenuFAR} also shows that a $1000$-hours observation with SKA should be able to not only detect the signal for the plotted models on a large range of wavenumbers, but should be sensitive enough to discriminate between the different Lyman-$\alpha$ coupling modellings. Note that a $100$ h survey (the medium survey in \citet{Koopmans15}), with a $\times 10$ higher thermal noise for the power spectrum, would already detect the signal in these favourable cases.

\section{Conclusions}

The modelling of the Lyman-$\alpha$ coupling of the hydrogen spin temperature to the gas kinetic temperature has a long history. While the theory has been established more than fifty years ago \citep{Wouthuysen52,Field58}, the necessity of computing the local spectrum around the Lyman-$\alpha$ line has led, as least initially to drastic simplifications. Initially, the coupling was simply assumed to saturate very fast and thus predictions during the Cosmic Dawn, when the spatial fluctuations of this
coupling are a dominant contribution to the 21-signal, were unreliable. Then, both semi-numerical methods and full radiative transfer codes where developed to compute the coupling and produce more accurate predictions for the 21-cm signal during the Cosmic Dawn. Recently, \cite{Reis21} improved the modelling of semi-numerical codes to include the effect of wing scatterings. 

In this work, we performed a careful comparison of the results from the semi-numerical method and a full radiative transfer code. We find that, ignoring the role of gas peculiar velocity, the two methods agree to a good level, even though the semi-numerical method assumes a propagation in a homogeneous universe. However, we find that when the Doppler effect from gas velocities is included (as it is by default) in the full radiative transfer code, the resulting 21-cm signal power spectrum is modified by up to a factor of $\sim 2$ at some scales and redshifts that depend on other model parameters. We presented some theoretical estimates of the expected amplitude of the effect of velocities on the local Lyman-$\alpha$ flux. We showed that the effect of velocities can be analysed by distinguishing two contributions. The first occurs while the photons are still in the wing of the Lyman-$\alpha$ line; it mainly changes the location where the photons will redshift into the core. The second contribution occurs during the propagation in the core of the line; it changes the number of scatterings a photon will undergo before redshifting out of the core. Both effects tend to counter each other but do not necessarily balance out. Finally we showed that various 21-cm signals resulting from different levels of Lyman-$\alpha$ modelling, everything else being identical, can be distinguished with high significance by the SKA, while the same is true for the NenuFAR CD KSP only for "exotic" models including an additional radio background.

At this stage, it is unclear how the effect of velocities could be included in a fast semi-numerical modelling. We have obtained very good results by training a neural work to produce a correction to be applied to the output of SPINTER to reproduce the output of LICORICE. However, this network was trained with data from a specific model at a specific redshift (using half of the data cube for training and half for testing). To be confident with using such a network, it would need to be trained with a variety of models and different redshifts. That would require running many LICORICE simulations to build the training sample, with stringent constraints on the redshift interval averaging. That would be a challenge in terms of computing time. Furthermore, if such a training sample was available, a network could probably be trained not to compute a correction to SPINTER but probably as an emulator to LICORICE. It is not impossible that using a modified kernel in SPINTER, that would implement an average effective radial velocity profile (probably redshift dependent), could lead to an improved agreement. A theoretical framework remains to be formulated for such an approach to be more than just phenomenological. 

\begin{acknowledgements}
      Some of the numerical simulations in this work were performed using HPC resources from GENCI-CINES (grant 2018-A0050410557). LVEK acknowledges the financial support from the European Research Council (ERC) under the European Union’s Horizon 2020 research and innovation programme (Grant agreement No. 884760, "CoDEX”). RB acknowledges the Israel Science Foundation (grant No. 2359/20), the Vera Rubin Presidential Chair in Astronomy and the Packard Foundation.
\end{acknowledgements}

\bibliographystyle{aa}
\bibliography{myref}

\end{document}